\documentclass[eqsecnum,showpacs,superscriptaddress]{revtex4}

\newcommand{\beq}{\begin{equation}}
\newcommand{\eeq}{\end{equation}}

\begin{document}

%%%%%%%%%%%%%%%%%%%%%%%%%%%%%%
\title{Correlator Bank Detection of GW chirps.\protect\\
False-Alarm Probability, Template Density and Thresholds: \protect\\
Behind and Beyond the Minimal-Match Issue.}
%%%%%%%%%%%%%%%%%%%%%%%%%%%%%%
\author{R.P. Croce}
\affiliation{Wavesgroup, University of Sannio at Benevento, Italy}
\author{Th. Demma}
\affiliation{Wavesgroup, University of Sannio at Benevento, Italy}
\author{M. Longo}
\affiliation{DIIIE, University of Salerno, Italy}
\author{S. Marano}
\affiliation{DIIIE, University of Salerno, Italy}
\author{V. Matta}
\affiliation{DIIIE, University of Salerno, Italy}
\author{V. Pierro}
\affiliation{Wavesgroup, University of Sannio at Benevento, Italy}
\author{I.M. Pinto}
\affiliation{Wavesgroup, University of Sannio at Benevento, Italy}

%\thanks{}
%%%%%%%%%%%%%%%%%%%%%%%%%%%%%%
%%%%%%%%%%%%%%%%%%%%%%%%%%%%%%
\begin{abstract}
The general problem of computing the false-alarm rate
vs. detection-threshold relationship for a bank of correlators is addressed,
in the context of maximum-likelihood detection of gravitational waves,
with specific reference to chirps from coalescing binary systems. 
Accurate (lower-bound) approximants for the cumulative distribution
of the whole-bank supremum are deduced from a class of Bonferroni-type inequalities.
The asymptotic properties of the cumulative distribution are obtained,
in the limit where  the number of correlators goes to infinity. 
The validity of numerical simulations made on small-size banks is extended
to banks of any size, via a gaussian-correlation inequality.
The result is used to estimate the optimum template density,
yielding  the best tradeoff between  computational cost
and detection efficiency, in terms of undetected potentially observable
sources at a prescribed false-alarm level,
for the simplest case of Newtonian chirps. 
\end{abstract}
%%%%%%%%%%%%%%%%%%%%%%%%%%%%%%
\date{}
%***************************************************
\maketitle

\section{Introduction.}
\label{sec:Intro}
%***************************************************

In the next few years a number of optical-interferometric detectors
of gravitational waves will be in operation
\cite{TAMA300}-\cite{VIRGO},
supplying steady fluxes of data to be sieved in the search of
gravitational wave (henceforth GW) signatures of astrophysical origin.
Among expected signals, those of  {\em known} shape, 
except for the {\em values} of a set of parameters 
which embody the relevant information on the source physics, 
are specially valuable to perspective gravity wave astronomy \cite{GW_astronomy}.
The maximum-likelihood strategy for detecting signals
of known shape in  additive  stationary gaussian colored noise \cite{noise}
consists in projecting the data on a finite discrete set
of  expected waveforms (the so called {\em templates}), 
and using the largest projection $\Lambda$ for computing a detection statistic.
The detection statistic $\Lambda$ is a random variable. Letting
\beq
F_\Lambda(x)=\mbox{Prob}[\Lambda \leq x |\mbox{no signal}],
\label{eq:the_star}
\eeq
it is possible to determine  uniquely a {\em threshold} $\gamma$ 
such that:
\beq
\mbox{Prob}[\Lambda > \gamma|\mbox{no signal}] = 1-F_\Lambda(\gamma)= \alpha
\eeq
where $\alpha$ is a {\em prescribed} false-alarm probability.
The detection algorithm accordingly consists in comparing the
detection statistic $\Lambda$ to the threshold $\gamma$. 
Whenever $\Lambda > \gamma$ one declares that a signal has been observed at
the false-alarm level $\alpha$. 
The unknown source parameters are then
{\em estimated} from those of the template yielding the largest
projection.
Deriving an {\em exact} (or suitably approximate) representation 
of the distribution function $F_\Lambda(x)$   
is clearly a key problem for detector's efficiency optimization.
In this paper we present a thorough analysis  
and a possible working solution of  this problem, 
for the special relevant case of gravitational wave chirps 
from coalescing binary stars. 
In particular, we present the formal deduction of several
partial results anticipated in \cite{CQG1},\cite{CQG2}.
As a by-product, previous findings by Dhurandhar, Schutz,
Krolak and Mohanty \cite{Dhur_Schu}-\cite{Mohanty} 
are recovered and hopefully generalized.
The paper is organized as follows.
In Sect. \ref{sec:Templates}, gravitational wave chirp templates and correlators
are introduced, the minimal match is defined, and the statistical
covariance among different correlators is computed, 
for the special relevant case where no signal is present in the data.
This section is included to make the paper self-contained
and to fix the notation, and could perhaps be skipped by the expert Reader.
In Sect. \ref{sec:MM_vs_undetect}, the minimal-match issue is reviewed, 
and the interplay
between template density, 
detection threshold 
and detector's efficiency 
(fraction of detected events among those potentially observable) 
is discussed. 
The study of  (\ref{eq:the_star}) is undertaken in Sect. \ref{sec:Core}
Available results are first reviewed.
A number of accurate  
lower-bound (conservative) {\em approximants} of $F_\Lambda(x)$ are then
discussed. 
The abstract asymptotic properties of $F_\Lambda(x)$ 
in the limit where the number of correlators $N \rightarrow \infty$
are next highlighted.
It is finally shown that one can capitalize on a classical
inequality to extend the validity of numerical simulations made on moderate-size 
template banks to banks of arbitrary (large) size.
In Sect. \ref{sec:MM_again} the minimal-match issue
is reconsidered, in the light of results obtained in previous sections.
The main conclusion  is that, 
in contrast to common belief,
increasing the template density in parameter space 
beyond a certain limiting value 
does {\em not} improve the detector's performance
(fraction of detected events among those potentially observable)
in any significant fashion. 
It is pointed out that in order to estimate 
the above limiting value, 
the accuracy of the no-signal cumulative distribution (\ref{eq:the_star})
is a key ingredient.
Conclusions and recommendations follow under Sect. \ref{sec:Conclu}.

The core results of this paper (especially those in Sect. \ref{sec:Core})
rely  on a number of rather technical arguments, which are
collected in Appendix A and B.

%%%%%%%%%%%%%%%%%%%%%%%%%
\section{Chirp Templates and Correlators}
\label{sec:Templates}
%%%%%%%%%%%%%%%%%%%%%%%%%

The (spectral) projection of the  data $A(f)$ on a template $T(f)$ 
is the (antenna noise-weighted) scalar product:
\beq
\langle A,T \rangle
=2 \mbox{Re}
\left\{ 
\int_{f_i}^{f_s}\frac{df}{\Pi(f)}A(f)T^*(f)
\right\},~~~
\label{eq:correlator} 
\eeq 
also called the {\em correlator} between $A$ and $T$.
In (\ref{eq:correlator}), $A(f)\!=\!S(f)\!+\!N(f)$, 
where $S(f)$ is a (possibly null) signal,
$N(f)$ is a realization of the antenna noise,
$(f_i,f_s)$ is the useful antenna spectral window,
$\Pi(f)$ is the antenna noise power spectral density, and $*$
denotes complex conjugation.\\
The stationary-phase asymptotic representation of 
a post-newtonian chirp waveform can be written \cite{PN_phase}:
\beq
H(f,\phi,\tau,\vec{\xi})=H_0 f^{-7/6}
\exp\left\{j\left[ 2\pi f \tau-\phi+\psi(f,\vec{\xi})\right]\right\},
\label{eq:chirp}
\eeq
where $H_0$ is a (complex, unknown) constant, 
depending on the source position and (relative) antenna orientation, 
$\tau$ is the nominal coalescency time \cite{nominal_Tc}, 
$\phi$ is the phase at $t=\tau$ 
and $\psi(f,\vec{\xi})$ is a function of  
the {\em intrinsic} source parameters only
(i.e., the binary companion masses and spin parameters), 
represented  by the vector $\vec{\xi}$.
Equation (\ref{eq:chirp}) is used to model the sought signal 
as well as to construct the templates.
Suffixes $S$ and $T$ will be used, when needed, 
to identify the signal and template, respectively. 
Templates are usually constructed with unit norm, viz.:
\beq
H_{0_T}=\left[2
\int_{f_i}^{f_s} f^{-7/3}\frac{df}{\Pi(f)}
\right]^{-1/2}
\Longleftrightarrow
||T||=
\langle T,T \rangle^{1/2} = 1.
\label{eq:unit_norm}
\eeq
The maximum-likelihood detection statistic is 
(any monotonically increasing function of):
\beq
\Lambda = \sup_{\vec{\xi}_T} \sup_{\tau_T} \sup_{\phi_T} \langle A,T \rangle,
\label{eq:det_stat0}
\eeq
where the supremums are taken on the  set of available templates.\\
The correlators (\ref{eq:correlator}) can be readily maximised 
w.r.t. the irrelevant parameter $\phi_{T}$, yielding:
\beq
c[A,T]=\sup_{\phi_{T}} \langle A,T \rangle =
\frac{
\displaystyle{
\left|
2 \int_{f_i}^{f_s}
A(f) f^{-7/6} 
e^{
-j
\left[
2\pi f \tau_T
+\psi(f,\vec{\xi}_T)
\right]
}
\frac{df}{\Pi(f)}
\right|
}}
{
\displaystyle{
\left[2
\int_{f_i}^{f_s} f^{-7/3}\frac{df}{\Pi(f)}
\right]^{1/2}
}}.
\label{eq:noncoh_correl}
\eeq
The r.h.s. of (\ref{eq:noncoh_correl}) is called the {\em  noncoherent correlator}.
In view of eq. (\ref{eq:noncoh_correl}) all templates can be generated with $\phi_{T}=0$.
The integral in  (\ref{eq:correlator})  is then recognized 
(but for inessential factors) as the $f \rightarrow \tau_{T}$ 
Fourier transform of the function:
\beq
K_A(f,\vec{\xi}_T)=
\left\{
\begin{array}{l}
2\displaystyle{\frac{A(f)T^*(f,0,0,\vec{\xi}_T)}{\Pi(f)}},~~~f \in (f_i, f_s)\\
0,~~~~~~~~~~~~~~~~~~~~~~~~~\mbox{elsewhere}.
\end{array}
\right.
\label{eq:four_kern}
\eeq
One is therefore led to construct a family of {\em reduced templates}
with $\phi_{T}=\tau_{T}=0$, parameterized in $\vec{\xi}_T$.
Further maximisation of the noncoherent correlator 
(\ref{eq:noncoh_correl}) w.r.t. $\tau_{T}$ 
can thus be accomplished by taking the largest absolute value 
of the $f \rightarrow \tau_T$ Fourier transform of  (\ref{eq:four_kern}). 
The resulting quantity 
\beq
C[A,T] = 
\sup_{\tau_{T}} 
c[A,T] =
\sup_{\tau_{T}} 
\sup_{\phi_{T}}
\langle A,T \rangle= 
\sup_{\tau} \left| {\cal F}_{f \rightarrow \tau}[K_A(f,\vec{\xi}_T)] \right|
\label{eq:red_correl}
\eeq
is called the {\em reduced} noncoherent correlator, 
and is a function of the {\em intrinsic} source and template parameters only.
The detection statistic can thus be  written
\beq
\Lambda = \sup_{\vec{\xi}_T} C[A,T]. 
\label{eq:det_stat}
\eeq

%---------------------------------------------------
\subsection{The Minimal-Match Prescription}
\label{subsec:MinMatch}
%---------------------------------------------------

The similarity (matching)  between a signal $S(f)$ 
and a (non reduced) template $T(f)$ is measured by their {\em overlap},
\beq
{\cal O}[S,T]=
\frac{\langle S,T \rangle}{|| S || \cdot || T ||} \leq 1
\label{eq:overlap}
\eeq
where the norm is defined in eq. (\ref{eq:unit_norm}).\\
Given a (discrete) family of templates ${\cal L}$,
and a set  ${\cal S}$  of admissible signals, the quantity
\beq
FF({\cal L},{\cal S})=
\inf_{S \in {\cal S}}
\sup_{T \in {\cal L}}
{\cal O}[S,T]
\label{eq:FF_set}
\eeq
is called the fitting-factor  between ${\cal L}$ and ${\cal S}$  \cite{Apostolatos}.

The probability of detecting 
(at any {\em fixed} false-alarm level)
a known signal $S(f)$  
hidden in the noisy (spectral) data $A(f)$
using $\langle A,T \rangle$ as a detection statistic
is an increasing function \cite{Apostolatos} of the overlap ${\cal O}[S,T]$.
Thus, given ${\cal S}$,  the (discrete, non-reduced) template family ${\cal L}$ 
is designed in such a way that 
\beq
FF({\cal L},{\cal S}) \geq \Gamma,
\label{eq:FF_bnd}
\eeq
where the physical meaning of $\Gamma$ will be discussed in Sect. \ref{sec:MM_vs_undetect}.
It is convenient to define the function
\beq
M[S,T]=
\sup_{\tau_{T}}
\sup_{\phi_{T}}
{\cal O}[S,T]
\label{eq:match}
\eeq
called the match between $S$ and $T$  \cite{Owen},  
and to rephrase (\ref{eq:FF_bnd}) as follows:
\beq
\forall S \in {\cal S},~\exists T \in {\cal L}:~
M[S,T] \geq \Gamma.
\label{eq:min_match}
\eeq
The quantity $\Gamma$ is referred to as the {\em minimal match}.

The match can be readily computed in terms of the {\em reduced}
templates and waveforms:
\beq
M[S,T]=
\left[
\int_{f_i}^{f_s}
f^{-7/3} 
\frac{df}{\Pi(f)}
\right]^{-1}
\sup_{\tau} 
\left| {\cal F}_{f \rightarrow \tau}[K_S(f,\vec{\xi}_T)] \right|,
\eeq
where, from eq. (\ref{eq:chirp})
\beq
K_S(f,\vec{\xi}_T)=
\left\{
\begin{array}{l}
\displaystyle{
\frac{f^{-7/3} 
e^{j\left[ \psi(f,\vec{\xi}_S)\!-\!\psi(f,\vec{\xi}_T)\right]}
}
{\Pi(f)}
},~~~f \in (f_i, f_s)\\
0,~~~~~~~~~~~~~~~~~~~~~~~~~~~~~~~~~\mbox{elsewhere}.
\end{array}
\right.
\label{eq:skern}
\eeq
More or less obviously,
\beq
\forall~\vec{\xi}_S, \vec{\xi}_T,~~~0 < M[S,T] \leq 1,
\eeq
the equality holding  iff $S=T$.

%---------------------------------------------------
\subsection{The (Reduced) Template Lattice}
\label{subsec:Lattice}
%---------------------------------------------------

We shall assume the reduced waveforms/templates to be parameterized 
in such a way  that  $\psi(f,\vec{\xi}_S)-\psi(f,\vec{\xi}_T)$ 
is a function of the {\em difference}  $\vec{\xi}_S-\vec{\xi}_T$ only \cite{Tag_Tan}.
Under this assumption, to leading order in the (small) difference $\vec{\xi}_S-\vec{\xi}_T$,
the match can  be cast in the form \cite{Owen}:
\beq
M[S,T] \approx 1-(\vec{\xi}_S  -  \vec{\xi}_T)^2
\label{eq:match_2}
\eeq
and the mimimal-match condition becomes:
\beq
\forall \vec{\xi}_S \in {\cal S}',~
\exists \vec{\xi}_T \in {\cal L}':~
(\vec{\xi}_S-\vec{\xi}_T)^2  \leq 1-\Gamma,
\label{eq:min_match2}
\eeq
where ${\cal S}' \subset{\cal S}$ and ${\cal L}' \subset{\cal L}$ are
the  sub-manifolds corresponding to $\phi=\tau=0$.
The ball 
\beq
\Sigma_k=\{\vec{\xi}_S : (\vec{\xi}_S-\vec{\xi}_{k})^2  \leq 1\!-\!\Gamma\}
\label{eq:span}
\eeq  
is called the {\em span} of  $T_k=T(\vec{\xi}_k)$  
at the  minimal-match level $\Gamma$,
and the minimal-match prescription is further rephrased as:
\beq
\bigcup_{k}  \Sigma_k \supseteq {\cal S}'.
\label{eq:min_match3}
\eeq
In the simplest case, the $\vec{\xi}_k$ can be chosen to form
a cubical lattice \cite{PierPin5}. 
Then, the distance between nearest-neighbour reduced templates 
(cubical-cell sidelength) is 
\beq
\Delta=2\left(\frac{1-\Gamma}{P}\right)^{1/2}
\label{eq:lattice_spacing}
\eeq
where $P$ is the dimension of $\vec{\xi}$.
Hence, the template density is $\Delta^{-P}$, and the template
number $N_{\Delta}$  is of the order of the volume of ${\cal S}'$ 
divided by the elementary cell volume $\Delta^P$.
 
%---------------------------------------------------
\subsection{The Pure-Noise Case Correlator Covariance Matrix}
\label{subsec:CovMat}
%---------------------------------------------------
In deducing the (no-signal) cumulative distribution of the whole-bank supremum it is 
expedient to deal with the following {\em squared} quantities:
\beq
Z=\frac{c^2}{2},~~~{\cal Z}=\frac{C^2}{2},~~~X=\frac{\Lambda^2}{2},
\label{eq:squared_cont}
\eeq
in view of the exponential (no-signal) distribution of the random variable $Z$.
The sought transformation does not affect detector performances, $X$ being an {\em equivalent} sufficient detection statistic. Equations (\ref{eq:squared_cont}) will be used systematically throughout.
Letting
\[
Y(\tau,\vec{\xi})=:
\frac{
\displaystyle{
\int_{f_i}^{f_s}
N(f) f^{-7/6} 
e^{
-j
\left[
2\pi f \tau
+\psi(f,\vec{\xi})
\right]}
\frac{df}{\Pi(f)}
}
}
{
\displaystyle{
\left[
\int_{f_i}^{f_s} f^{-7/3}\frac{df}{\Pi(f)}
\right]^{1/2}
}}=
\]
\beq
=\left[
\int_{f_i}^{f_s} f^{-7/3}\frac{df}{\Pi(f)}
\right]^{-1/2}
{\cal F}_{f \rightarrow \tau}[K_N(f,\vec{\xi})],
\label{eq:complex_Y}
\eeq 
where
\beq
K_N(f,\vec{\xi})=
\left\{
\begin{array}{l}
\displaystyle{
\frac{
N(f)
f^{-7/6} 
e^{
-j
\psi(f,\vec{\xi})
}
}{\Pi(f)}},~~~f \in (f_i, f_s)\\
0,~~~~~~~~~~~~~~~~~~~~~~~~~\mbox{elsewhere},
\end{array}
\right.
\label{eq:random_fun}
\eeq
one readily gets,
in view of eq.s (\ref{eq:noncoh_correl}) and (\ref{eq:squared_cont}):
\beq
Z=|Y(\tau,\vec{\xi})|^2,
~~~{\cal Z}=\sup_{\tau} \left| Y(\tau,\vec{\xi})  \right|^2,
~~~X=\sup_{\vec{\xi}} \sup_{\tau} \left| Y(\tau,\vec{\xi}) \right|^2.
\eeq
The noise being assumed as stationary zero-mean colored gaussian,
$Y(\tau,\vec{\xi})$ is a zero-average {\em complex gaussian random field}.
It is therefore completely characterized by its second-order 
joint expectation value (covariance), viz.:
\beq
E[Y(\tau,\vec{\xi})Y^*(\tau',\vec{\xi}')]=
\frac{\displaystyle{
\int_{f_i}^{f_s}
f^{-7/3} 
e^{-j
\left\{
2\pi f (\tau-\tau') +
\left[
\psi(f,\vec{\xi})-
\psi(f,\vec{\xi}')
\right]
\right\}
}
\frac{df}{\Pi(f)}
}
}{
\displaystyle{
\int_{f_i}^{f_s}
f^{-7/3} 
\frac{df}{\Pi(f)}
}}.
%=:\Sigma(\Delta \tau, \vec{\xi},\vec{\xi}')
\label{eq:cov_mat}
\eeq
It is readily seen that:
\beq
\left|
E[Y(\tau,\vec{\xi})Y^*(\tau',\vec{\xi}')]
\right|=
\frac{c[T,T']}
{||T||\cdot||T'||}. 
\label{eq:quaqqua}
\eeq  
 
%---------------------------------------------------
\subsection{Discrete Implementation}
\label{subsec:Discrete}
%---------------------------------------------------

Let $N_{\Delta}$ the total number of reduced templates used to cover ${\cal S}'$.
Each reduced template yields as many (non-coherent) correlators
as the number of time bins in the discrete Fourier transform (henceforth DFT \cite{DFT})
used to compute numerically eq. (\ref{eq:noncoh_correl}).
The above number will be denoted by $N_{\Theta}$. 
The detection statistic is thus:
\beq
\Lambda=
\max_{k \in [1,N_{\Delta}]}
\max_{h \in [1,N_{\Theta}]}
\left|
DFT_h[K_A(f,\vec{\xi}_k)]
\right|,~~
\label{eq:discr_det_stat}
\eeq
where the integers $h$ and $k$ span the DFT output time bins,
denoted as $DFT_h[\cdot]$, and the (discrete) reduced template family, respectively.

The match between a signal  $S$ and a template $T$ is computed as
\beq
M[S,T]=
\max_{h \in [1,N_{\Theta}]}
\frac{
\left|
DFT_h[K_S(f,\vec{\xi}_k)]
\right|}
{\displaystyle{
\left|
\int_{f_i}^{f_s}
f^{-7/3} 
\frac{df}{\Pi(f)}
\right|
}}.
\label{eq:match_d}
\eeq
As an effect of time-discretization, $M[S,S]$ computed with (\ref{eq:match_d})
can be less than one. 
The DFT time step  (and length)  should be accordingly chosen in such a way that 
\beq
\forall S \in {\cal S}', ~~~1-M[S,S] \ll 1-\Gamma.
\label{eq:tau_t_sampling}
\eeq
We define the following zero-average gaussian random suite, 
which is the discrete counterpart of eq. (\ref{eq:complex_Y})
\beq
Y_{h,k}=
\left[
\int_{f_i}^{f_s} f^{-7/3}\frac{df}{\Pi(f)}
\right]^{-1/2}
DFT_h[K_N(f,\vec{\xi}_k)],~~
h=1,2,\dots,N_{\Theta},~~k=1,2,\dots,N_{\Delta},
\label{eq:Y_DFT}
\eeq
which is completely characterized by the covariance matrix
\beq
\sigma(k-h,\vec{\xi}_n-\vec{\xi}_m)=E
\left[
Y_{h,m} Y^*_{k,n}
\right].
\eeq
We note that $\sigma(0,\vec{0})=1$ by construction, and further define
\beq
\rho_M = \max |\sigma(q,\vec{\xi}_n-\vec{\xi}_m)|,
~~~(q,\vec{\xi}_n-\vec{\xi}_m)\neq (0,\vec{0}).
\eeq
Note that, in view of eq. (\ref{eq:tau_t_sampling}) one has
\beq
\rho_M=|\sigma(1,\vec{0})|
\eeq 
which means that the largest (absolute) covariance always occurs between correlators
differing only in $\tau$ by a single DFT time bin.

%%%%%%%%%%%%%%%%%%%%%%%%%%%%%%%%
\section{Minimal Match vs. Undetected Observable Sources}
\label{sec:MM_vs_undetect}
%%%%%%%%%%%%%%%%%%%%%%%%%%%%%%%%

The minimal match
is more or less obviously related to
the {\em fraction} of potentially observable sources 
which would be undetected as an effect of template mismatch.
It was first argued in \cite{Apostolatos} that
(under the simplest assumption of isotropically 
and homogeneously distributed sources) this
latter would be simply $(1-\Gamma^3)$.

In order to gain insight into this issue, it is expedient
to work in terms of the detection probability $P_d$.
Accordingly, let 
\beq
P_d=1-\beta=1-Q[\gamma-E(\Lambda)]
\label{eq:det_prob}
\eeq
where $Q(\cdot)$ is an unknown CDF, whose density can be nonetheless
safely assumed as being unimodal and center-symmetric, at least
for meaningful values of the signal-to-noise ratio.
In the worst case of {\em minimal} match,
\beq
E(\Lambda) \approx \Gamma d,
\eeq
where $d$ is the signal deflection (signal-to-noise ratio), viz.:
\beq
d:=\left|2 
\int_{f_i}^{f_s}
\frac{S(f)S^*(f)}{\Pi(f)} df
\right|^{1/2},
\label{eq:deflec}
\eeq
and we assume $d \gg 1$.
Let the {\em observable} sources be those for which $P_d \geq 0.5$.             
Under the above assumptions for $Q(\cdot)$, these correspond to:
\beq
d \geq \frac{\gamma}{\Gamma}.
\label{eq:defletto}
\eeq 
The deflection $d$ on the other hand can be written
\beq
d=\frac{K}{R}
\eeq
where $R$ is the source distance, and $K$ is a constant 
depending on the antenna and source  features, as well
as their mutual orientation.
Equation (\ref{eq:defletto}) allows to define 
the radius $R$ of the antenna-centered sphere
which contains all observable sources of a given class
(seen by the same antenna, sharing the same features and orientation), viz.:
\beq
R=\frac{K\Gamma}{\gamma}.
\eeq
Hence the number of {\em observable} sources
in the given class, 
under the most obvious hypothesis of uniform 
(homogeneous and isotropic) spatial source density $\rho_s$ is:
\beq
N_s=\frac{4}{3}\pi\rho_s
\left[
\frac{K\Gamma}{\gamma}
\right]^3.
\label{eq:no_of_sources}
\eeq
Letting $\Gamma_{max}$, $N_{max}$ the largest achievable
minimal match, and the related number of templates,
and denoting as $N_s^{(max)}$ the corresponding value of $N_s$,
one has from (\ref{eq:no_of_sources}):
\beq
\eta=\frac{N_s}{N_s^{(max)}}=
\left(\frac{\Gamma}{\Gamma_{max}}\right)^3
\left(\frac{\gamma(\alpha,N_{max})}{\gamma(\alpha,N)}\right)^3
\label{eq:eta}
\eeq
where the dependence of $\gamma$ from $\alpha$ and $N$ 
(and hence on $\Gamma$ itself) 
has been written explicitly. The quantity $1-\eta$ represents 
the fraction of {\em observable} sources which would
be lost due to poor minimal match ($\Gamma < \Gamma_{max}$).
Note that the ratio $\eta$ does {\em not} depend on $K$.\\
Neglecting the dependence of $\gamma$ on $N$,
the second factor on the r.h.s. of (\ref{eq:eta}) cancels, 
and we are left with
\beq
\eta=\left(\frac{\Gamma}{\Gamma_{max}}\right)^3:=\eta_0,
\label{eq:eta0}
\eeq
which does {\em not} depend on $\alpha$.
Under the same assumption, $N_s$ in (\ref{eq:no_of_sources}) 
increases monotically with  $\Gamma$, 
attaining its supremum $N_s^{(max)}$ at $\Gamma_{max}=1$,
yielding $\eta_0=\Gamma^3$.
This is, in essence, the simplest argument 
introduced in \cite{Apostolatos}.\\
This argument conceals an obvious naivety insofar
as it neglects the dependence of the detection threshold $\gamma$ 
corresponding to a given false-alarm probability $\alpha$
on the total number of templates $N=N_\Theta N_\Delta$,
and hence on $\Gamma$ itself.\\

In order to refine this reasoning one should fully take into account
the dependence of $\gamma$ on $N$, which in turn is affected by
the covariance among the correlators.
We shall postpone the relevant results 
of such a refined analysis 
until the end of Sect. \ref{sec:MM_again},
after having introduced the needed 
no-signal CDF of the detection statistic $\Lambda$, which is
the subject of the next section.

%%%%%%%%%%%%%%%%%%%%%%%%%%%%%%%%%
\section{No-Signal  Whole-Bank Statistics}
\label{sec:Core}
%%%%%%%%%%%%%%%%%%%%%%%%%%%%%%%%%

In order to provide a "working" solution to the problem of
computing the whole-bank false-alarm probability, we should be able
in principle:
\begin{itemize}
\item{to provide a hopefully decent {\em approximant} of $F_X$;}
\item{to guarantee that the above approximant is otherwise a {\em lower bound}
for the {\em true} CDF, so that one makes a conservative error
while using it to compute the threshold 
corresponding to a given false-alarm probability;}
\item{to formulate a criterion whereby the validity of simulations made on relatively
small-sized banks could be extended to huge (real-world) banks, for which direct numerical
simulations are either unwieldy or unfeasible.}
\end{itemize}
To make the dependence of  $F_X(x)$ 
on the number $N$ of templates explicit, we shall denote it in this section by
\beq
F^{(N)}_X(x)=\mbox{Prob}[X \leq x | \mbox{no signal}].
\label{eq:cumulative}
\eeq

%%%%%%%%%%%%%%%%%%%%%%%%%%%%%%%%%%%
\subsection{Background}
\label{subsec:Bckgrnd}
%%%%%%%%%%%%%%%%%%%%%%%%%%%%%%%%%%%

In the ideal case where the correlators are
statistically independent, one has simply:
\beq
F^{(N)}_X(x)=[1-\exp(-x)]^{N}=:\Phi^{(N)}_X(x).
\label{eq:independent}
\eeq
In all practical cases,  
the covariance among the correlators
is nonzero, and hence the independent-correlator formula (\ref{eq:independent}) 
is poorly accurate. However, in Appendix A we prove that, 
for any finite $N$  and any $x$,  
the independent-correlator approximation $\Phi_X^{(N)}$
is  a {\em lower bound} for $F_X^{(N)}$, viz.:
\beq
F_X^{(N)} \geq \Phi_X^{(N)},
\label{eq:Bound_indep}
\eeq
thus implying
a {\em conservative} error (for any fixed threshold
it gives a lower false alarm rate).

It was first suggested by Dhurandhar and Schutz \cite{Dhur_Schu}
that  eq. (\ref{eq:independent}) can be made more accurate
after {\em reducing}  $N$ by a suitable factor  $\epsilon \in [0,1[$.
Extensive numerical simulations supporting this suggestion 
were subsequently reported in \cite{Dhur_Mo} 
by Dhurandhar and Mohanty  \cite{Dhur_Mo_detail}.
In Jaranowski et al. \cite{Jaranowski_et_Al} it was heuristically
argued that
\beq
\epsilon \approx \frac{\mu(\eta_T)}{\mu(\eta_c)}
\eeq
where $\mu(\eta_T)$, $\mu(\eta_c)$ are the  measures
of the {\em span} 
%\cite{span} 
and {\em correlation neighbourhood} 
\cite{corr_neigh} of a template, respectively \cite{Jaranowski_et_Al},
which are both functions of the prescribed minimal match $\Gamma$.
In \cite{Mohanty}  Mohanty introduced an elegant formal argument 
whereafter an approximate closed-form expression for $\epsilon$ 
could be derived \cite{Moh_clue}.
Mohanty's argument is based  on the following key assumptions,
valid for {\em suitably large}  values of the threshold:
\begin{itemize}
\item[i)]{
that {\em at most  two} (squared) correlators, say $Z$, $Z'$ can simultaneously exceed 
the threshold $x$, 
and 
}
\item[ii)]{
that the corresponding templates can only be nearest-neighbours,
featuring the largest (absolute) covariance.
As already noted, the largest (absolute) covariance $\rho_M$ occurs
between $Y_{i,j}$ and $Y_{i+1,j}$, so that Mohanty's argument
applies naturally to the cumulative distribution of each and
any reduced noncoherent correlator ${\cal Z}$. }
\end{itemize}
Hence \cite{Markov},
\beq
F^{(N_{\Theta})}_{\cal Z}(x) \approx 1-N_{\Theta}~\mbox{Prob}[Z_{i,j} > x|\mbox{no signal}]+
(N_{\Theta}-1)~\mbox{Prob}[Z_{i,j} > x, Z_{i+1,j} > x |\mbox{no signal}].
\label{eq:Moh_formula}
\eeq
This can be used to obtain the mentioned $\epsilon$ factor
%to be used in (\ref{eq:independent}) 
by comparison with
\beq
F^{(N_{\Theta})}_{\cal Z}(x) \approx 
[1-\exp(-x)]^{\epsilon N_{\Theta}} \approx 1-\epsilon N_{\Theta} \exp(-x),
\label{eq:binomial}
\eeq
(binomial approximation, large thresholds) yielding:
\beq
\epsilon \approx 1-(1-N_{\Theta}^{-1})\mbox{Prob}[Z_{i+1,j}>x|Z_{i,j}>x,\mbox{no signal}].
\label{eq:eps_moh}
\eeq

The above quoted results do not allow to make any rigorous statement 
insofar as the behaviour of $F_X^{(N)}$ is concerned for {\em large} $N$
and/or large thresholds (beyond values for which numerical checking is feasible).
In addition it is not clear whether using (\ref{eq:Moh_formula})
gives a conservative estimate of the threshold.

%%%%%%%%%%%%%%%%%%%%%%%%%%%%%%%%%%%
\subsection{Mohanty's formula as a lower bound}
\label{subsec:Mohanty}
%%%%%%%%%%%%%%%%%%%%%%%%%%%%%%%%%%%
In this section we derive Mohanty's formula following a different route,
whose validity does {\em not} rest on any of the {\em ad hoc} hypotheses made in \cite{Mohanty}. We shall further prove that Mohanty's formula is in fact a CDF lower bound. 

We start from the following {\em Bonferroni-type} inequality \cite{Bonfer}:
\[
\mbox{Prob}\left[\bigcup_{i=1}^{N} E_i\right]=
\mbox{Prob}\left[\bigcup_{i=1}^{N-1} \left(E_i-E_i\cap E_{i+1}\right)\cup
E_{N}\right]\le
\]
\beq
\le
\sum_{i=1}^{N-1}\mbox{Prob}\left[E_i-E_i\cap E_{i+1}\right]+\mbox{Prob}[E_{N}]
=
\sum_{i=1}^N\mbox{Prob}[E_i]-
\sum_{i=1}^{N-1}\mbox{Prob}[E_i\cap E_{i+1}],
\label{eq:Bonfercaz}
\eeq
which holds for arbitrary {\em events} $\{E_i\}_{i=1}^N$.
Equation (\ref{eq:Bonfercaz}) can be used to derive several lower bounds for the cumulative distribution of the {\em reduced} noncoherent correlator ${\cal Z}$. 
Indeed, letting $E_i=\{Z_{i,j}>x\}$,$~~i=1,2,\dots,N_\Theta$, where $j\in(1,2,\dots,N_\Delta)$ is fixed, one gets:
\[
F_{\cal Z}^{(N_\Theta)}(x)=1-\mbox{Prob}
[\max_{i=1,\dots,N_\Theta} Z_{i,j}>x|\mbox{no signal}] 
\geq 1 -
\sum_{i=1}^{N_\Theta} \mbox{Prob}[~Z_{i,j} > x|\mbox{no signal}~] +
\]
\beq
+\sum_{i=1}^{N_\Theta-1} \mbox{Prob}[~Z_{i,j} > x,Z_{i+1,j} > x|\mbox{no signal}~].
\label{eq:IE2}
\eeq
Equation (\ref{eq:IE2}) yields different lower bounds, 
depending on the way the templates are indexed. The tightest bound 
is obtained whenever the indexing is such that 
{\em consecutive} templates in (\ref{eq:IE2})
exhibit the {\em largest} covariance.
In view of eq. (\ref{eq:tau_t_sampling}) this occurs 
invariably between templates corresponding to the
same {\em intrinsic} parameters, 
and differing only  by a single time-discretization
bin in $\tau$ \cite{hierarchical}.

Taking into account the stationarity of the sequence $Z_{i,j}$, 
eq. (\ref{eq:IE2}) can be cast in the form:
\beq
F_{\cal Z}^{(N_\Theta)}(x) \geq 1 -
N_\Theta \mbox{Prob}[~Z_{i,j} > x|\mbox{no signal}~] +
(N_\Theta-1) \mbox{Prob}[~Z_{i,j} > x,Z_{i+1,j} > x|\mbox{no signal}~].
\label{eq:IE3}
\eeq 

Both probability terms in (\ref{eq:IE3}) 
can be evaluated explicitly, as shown in detail e.g.
in \cite{Levine}, yielding:
\beq
\mbox{Prob}[Z_{i,j} > x|\mbox{no signal}]=\exp(-x),
\label{eq:sum1}
\eeq
and
\beq
\mbox{Prob}[Z_{i,j} > x,Z_{i+1,j} > x|\mbox{no signal}]
=(1-\rho_M^2)\sum_{k=0}^{\infty}
\frac{\rho_M^{2k}{\mathit\Gamma}_E^2[k+1,(1-\rho_M^2)^{-1}x]}{{\mathit\Gamma}_E^2(k+1,1)},
\label{eq:sum2}
\eeq
where ${\mathit\Gamma}_E(\cdot,\cdot)$ 
is the incomplete Euler gamma function \cite{Prudni}.
Equation (\ref{eq:IE3}) can be conveniently written in a more compact form as:
\beq
F_{\cal Z}^{(N_\Theta)}(x) \geq 1 -
N_\Theta \exp(-x)\epsilon(N_\Theta,x,\rho_M),
\label{eq:compact}
\eeq
where:
\beq
\epsilon(N_\Theta,x,\rho_M)=
1-(1-N_\Theta^{-1})\exp(x)(1-\rho_M^2)
\sum_{k=0}^{\infty}
\frac{\rho_M^{2k}{\mathit\Gamma}_E^2[k+1,(1-\rho_M^2)^{-1}x]}{{\mathit\Gamma}_E^2(k+1,1)}.
\label{eq:epsilon1}
\eeq
Comparison of (\ref{eq:epsilon1}) with Mohanty's formula (\ref{eq:eps_moh}) 
reveals that this latter gives a {\em lower bound} \cite{lower_bnd}
of $F^{(N_\Theta)}_{\cal Z}$.
% for $N_{\Theta} \gg 1$.

In principle, inequality (\ref{eq:Bonfercaz}) can be used to derive lower bounds for the cumulative distribution of the whole-bank supremum $F^{(N)}_X$ as well.
For the sake of simplicity, we refer to the case of Newtonian templates,  featuring a single {\em intrinsic} parameter, namely the chirp mass.
Under this assumption, the $N=N_\Theta\dot N_\Delta$ correlators can be 
conveniently cast in lexicographic order:
\beq
Z_{i,j}\rightarrow Z_{i+(j-1)N_\Theta},~~~i=1,2,\dots,N_\Theta,~~~j=1,2,\dots,N_\Delta.
\label{eq:lessico}
\eeq
From (\ref{eq:lessico}) and Fig. 1, it is seen that there exist $N_\Delta-1$ consecutive correlator pairs whose covariances are negligible.
It turns out that the relevant number of correlator pairs exhibiting the maximum (absolute) covariance $\rho_M$ is $N_\Theta N_\Delta-(N_\Delta-1)-1$.
Accordingly, letting $E_i=\{Z_i>x\}$ in equation (\ref{eq:Bonfercaz}) gives:
\[
F^{(N)}_X(x)=1-\mbox{Prob}[\max_i Z_i>x|\mbox{no signal}]\ge
1-\sum_{i=1}^{N_\Theta N_\Delta}\mbox{Prob}[Z_i>x|\mbox{no signal}]+
\]
\beq
+
\sum_{i=1}^{N_\Theta N_\Delta-N_\Delta}
\mbox{Prob}[Z_i>x, Z_{i+1}>x|\mbox{no signal}],
\label{eq:corr1}
\eeq
where the meaning of $Z_i$ is obvious from (\ref{eq:lessico}).
Equation (\ref{eq:corr1}) is conveniently rephrased as:
\beq
F^{(N)}_X(x)\ge
1-N_\Theta N_\Delta \exp(-x)\epsilon(N_\Theta,x,\rho_M).
\label{eq:epsilon2}
\eeq

%%%%%%%%%%%%%%%%%%%%%%%%%%%%%%%%%%%
\subsection{Nested Mohanty's Formula}
\label{subsec:Nested}
%%%%%%%%%%%%%%%%%%%%%%%%%%%%%%%%%%%

Inequality (\ref{eq:Bonfercaz})
can be notably invoked {\em recursively} to provide  hopefully more accurate
lower bounds for the CDF of the supremum
of the {\em whole} bank of (squared noncoherent) correlators. 
Indeed, letting $E_j = \{{\cal Z}_j > x\}$, $j=1,2,\dots,N_\Delta$ 
in (\ref{eq:Bonfercaz}), we obtain:
\[
F_X^{(N)}(x) =1-\mbox{Prob}[\max_j {\cal Z}_j>x|\mbox{no signal}]\geq
\]
\beq
\geq
1-\sum_{j=1}^{N_\Delta} \mbox{Prob}[{\cal Z}_j>x|\mbox{no signal}]+
\sum_{j=1}^{N_\Delta-1}\mbox{Prob}[{\cal Z}_j>x,{\cal Z}_{j+1}>x|\mbox{no signal}].
\label{eq:moh_bis}
\eeq
Again, by statistical uniformity and stationarity,
\beq
F_X^{(N)}(x) \geq
1\!-\!N_{\Delta} ~ \mbox{Prob}[{\cal Z}_j >x|\mbox{no signal}]\cdot
\left\{
1\!-\!(1\!-\!N_{\Delta}^{-1}) ~ \mbox{Prob}[{\cal Z}_{j+1}\!>\!x|{\cal Z}_{j}\!>\!x, \mbox{no signal}]
\right\}, 
\label{eq:cazzill}
\eeq
for any $j$.
One can use eq. (\ref{eq:compact}) in (\ref{eq:cazzill}), to obtain
\beq
F_X^{(N)}(x)\geq 1\!-\!
N_{\Delta} N_{\Theta}\exp(-x)
\epsilon(N_\Theta,x,\rho_M)
\cdot
\left\{
1\!-\!(1\!-\!N_\Delta^{-1})~
\mbox{Prob}[ {\cal Z}_{j+1}\!>\!x| {\cal Z}_{j}\!>\!x, \mbox{no signal}]
\right\}.
\label{eq:cazzill2}
\eeq
There is {\em no} simple recipe for computing 
the conditional probability in (\ref{eq:cazzill2}). 
However one might speculate that
\beq
\mbox{Prob}[{\cal Z}_{j+1}>x|{\cal Z}_{j}>x, \mbox{no signal}]
\geq
\mbox{Prob}[Z_{i,j+1}>x|Z_{m,j}>x, \mbox{no signal}],
\label{eq:ansatz}
\eeq
for any $i \in (1,2,\dots,N_{\Theta})$  
and any $j \in (1,2,\dots,N_{\Delta}-1)$, 
where $m \in (1,2,\dots,N_{\Theta})$ 
is defined by the condition \cite{heuristic}
\beq
\rho_M'=
\left|
E[
Y_{i,j+1}Y^*_{m,j}
]
\right|
>
\left|
E[
Y_{i,j+1}Y^*_{h,j}
]
\right|,~~~\forall h\neq m.
\eeq
Under this assumption, equation (\ref{eq:cazzill2}) yields:
\beq
F_X^{(N)}(x)\geq 1-
N_{\Delta} N_{\Theta}\exp(-x)
\epsilon(N_\Theta,x,\rho_M)
\epsilon(N_\Delta,x, \rho_M').
\label{eq:nested}
\eeq
Equation  (\ref{eq:nested}) can be readily generalized  to post-newtonian 
intrinsic parameter spaces of dimension higher than one,
by iterating the above reasoning.
It is also recognized that eq. (\ref{eq:epsilon2}) corresponds to neglecting
the covariance between reduced templates
in (\ref{eq:nested}).

%%%%%%%%%%%%%%%%%%%%%%%%%%%%%%%%%%%
\subsection{Asymptotics}
\label{subsec:Asymp}
%%%%%%%%%%%%%%%%%%%%%%%%%%%%%%%%%%%

We shall now discuss 
the  asymptotic form 
of $F^{(N)}_X$ as  $N \rightarrow \infty$.
This question is properly dealt with
in the frame of extreme-value theory (henceforth EVT).
Note that the study of the above asymptotics via
numerical simulation is almost impossible 
(the tail behaviour requires a huge number of simulations)
unless some a-priory information is available.

The EVT philosophy can be understood intuitively
by noting that the cumulative distribution $F^{(N)}$ of the supremum
of a set of $N$ (equidistributed, unimodal) random variables
displays the following universal behaviour as $N \rightarrow \infty$:
i) $F^{(N)}$ becomes steeper and steeper, tending to a step-like function, 
and ii) the CDF inflection point moves forward in $x$ indefinitely.
As a result, in the limit $N \rightarrow \infty$, the CDF of the supremum is zero everywhere,
except at infinity, where its limiting value is one.
EVT is a general framework
for dealing with the above limiting process in a meaningful
way \cite{Castillo}, and relies on a suitable $N$-dependent
change of variable such that the location of the inflection point of  $F^{(N)}$
becomes independent of $N$. 
It turns out that for most {\em independent} random variables, 
it is possible to find two sequences $\{a_N\}$ and $\{b_N > 0\}$
such that
\beq
\lim_{N\rightarrow \infty}
F^{(N)}(a_N+b_N x)
\label{eq:EVT}
\eeq
exists, and can only be one among three {\em universal}
functions: the Frechet, Weibull and Gumbel distributions
\cite{Castillo}. 

Using EVT we can readily prove that as $N \rightarrow \infty$
the cumulative distribution (\ref{eq:independent}),
converges uniformly to the Gumbel distribution 
$G^{(N)}(x)$ \cite{Gumbel}.

In fact, letting $x=y+\log N$ in
(\ref{eq:independent}) one gets:
\beq
\Phi^{(N)}_X(y+\log N)= \left(1-\frac{e^{-y}}{N}\right)^{N}
\stackrel{N\rightarrow\infty}{\longrightarrow} e^{-e^{-y}}.
\label{eq:pre_gum}
\eeq
Switching back to the original variables, one  gets:
\beq
\Phi^{(N)}_X(x) 
\sim 
\exp[-N \exp(-x)] = G^{(N)}(x).
\label{eq:Gumbel}
\eeq
It is also readily proved that the Gumbel distribution $G^{(N)}(x)$ 
is an {\em upper bound}  for $\Phi^{(N)}_X(x)$, 
for any $N$ and $x$, as illustrated in Fig. 2.
From the inset in Fig. 2, one can appreciate
that $\sup_{x}[G^{(N)}(x)-\Phi^{(N)}_X(x)]~$ 
drops below $10^{-5}$ already at $N \sim 10^4$.

For the general realistic case where  the covariance among
the correlators is nonzero the existence of the $a_n$ and $b_n$ suites
in (\ref{eq:EVT}) is not guaranteed, nor the coincidence
with their zero-covariance counterparts is implied.
We  prove in Appendix A that
for {\em any} nonzero (fixed) template spacing,
i.e., for any {\em nonzero} covariance, 
$F^{(N)}_X(x)$ converges uniformly 
to the Gumbel distribution 
in the limit as  $N \rightarrow \infty$, viz.:
\beq
F^{(N)}_X(y+\log N)
\stackrel{N\rightarrow\infty}{\longrightarrow} e^{-e^{-y}}.
\eeq
The rate of convergence 
is more or less obviously {\em slower} the {\em smaller} the template spacing,
i.e., the {\em larger} the maximum (absolute) covariance.
This is illustrated in Fig. 3.

%%%%%%%%%%%%%%%%%%%%%%%%%%%%%%%%%%%
\subsection{Beyond Mohanty-type Formulas}
\label{subsec:Beyond}
%%%%%%%%%%%%%%%%%%%%%%%%%%%%%%%%%%%

Inspection reveals that the r.h.s.'s of Mohanty-type formulas 
(\ref{eq:compact}), (\ref{eq:epsilon2}), (\ref{eq:nested}) 
are bad candidates as CDF approximants.
As $x \rightarrow 0+$  they become  negative.
Even worse, they {\em do not} exhibit the correct (Gumbel) asymptotic
behaviour as $N \rightarrow \infty$.
Both these flaws can be remediated by introducing the 
{\em exponential} representations \cite{Rytov}
\beq
\left\{
\begin{array}{l}
\Psi^{(N_\Theta)}_{\cal Z}(x)=
\exp\left[-N_\Theta \epsilon(N_\Theta,x,\rho_M) \exp(-x) \right]\\
\\
\Psi^{(N)}_X(x)=
\exp\left[-N_\Theta N_\Delta \epsilon(N_\Theta,x,\rho_M) \exp(-x) \right]\\
\\
\tilde{\Psi}^{(N)}_X(x)=
\exp\left[-N_\Theta N_\Delta \epsilon(N_\Theta,x,\rho_M)\epsilon(N_\Delta,x,\rho_M') \exp(-x) \right]\\
\end{array}
\right.
\label{eq:Rytov_like}
\eeq
which are always non negative, have the correct (Gumbel)
asymptotic behaviour,  and reproduce Mohanty's formulas 
(\ref{eq:compact}), (\ref{eq:epsilon2}), (\ref{eq:nested}) for $x$ suitably large.
We shall prove soon that (\ref{eq:Rytov_like}) are in turn {\em lower bounds}
for the corresponding exact CDF's.
We note from (\ref{eq:Rytov_like}) that
\beq
Q_X^{(N)}(x)=\left[Q_X^{(M)}(x)\right]^{N/M},~~~\forall M < N,~ N \gg 1,
\label{eq:scaling}
\eeq
where $Q_X^{(N)}$ can be any of the distributions in (\ref{eq:Rytov_like}).
%%%%%%%%%%%%%%%%%%%%%%%%%
\subsection{Extrapolating Simulations: A Gaussian Correlation Inequality}
\label{subsec:Rytov}
%%%%%%%%%%%%%%%%%%%%%%%%%
The inequality
\beq
F^{(N)}_X(x) \geq [F^{(M)}_X(x)]^{N/M},~~~M \le N
\label{eq:GCI}
\eeq
where $F^{(M)}_X(x)$ is the CDF of the supremum
of  any {\em subset} of the whole bank containing  
$M \le N$ correlators
follows directly from the so called 
Gaussian Correlation Inequality (henceforth GCI) \cite{GCI}.
More or less obviously, the upper bound in (\ref{eq:GCI})
is attained  as $M \rightarrow N$.\\
Equation  (\ref{eq:GCI}) is a key to demonstrate 
that equations in (\ref{eq:Rytov_like}) are lower bounds for
the pertinent {\em true} cumulative distributions.\\
As an example, let us consider the case of $\tilde{\Psi}$.
Numerical simulations show that for {\em relatively small} banks,
\beq
F_X^{(M)}(x) \geq \tilde{\Psi}_X^{(M)}(x).
\eeq
On the other hand, in view of eq. (\ref{eq:scaling})  
\beq
\left[ \tilde{\Psi}_X^{(M)}(x)\right]^{N/M} = \tilde{\Psi}_X^{(N)}(x),~~~N \gg 1,
\eeq
and hence, from (\ref{eq:GCI}),
\beq
F_X^{(N)}(x) \geq \tilde{\Psi}_X^{(N)}(x),~~~\forall N \gg 1.
\eeq
The GCI can be used to extend the validity 
of {\em approximate} forms of $F_X$ 
tested on {\em relatively small} banks
to full-size banks, 
for which numerical simulations 
would be impossible or unwieldy.
%++++++++++++++++++++ SUBSEC: Accuracy & Thresholds +++++++++++++++++++++++++
\subsection{Accuracy and Thresholds}
\label{subsec:Thresh}
%++++++++++++++++++++++++++++++++++++++++++++++++++++++++++++++++++++++++++++
The numerical accuracy of the (lower-bound) approximants
(\ref{eq:Rytov_like}) is illustrated in Fig.s 4 and 5,
obtained from numerical simulations over $\sim 2 \cdot 10^4$ realizations.\\
Figure 4 refers to the CDF $F_{\cal Z}^{(N_\Theta)}$ of a single reduced correlator,
and shows that the approximation provided by $\Psi_{\cal Z}^{(N_\Theta)}$ is excellent.
Figure 5 refers to the CDF  of the supremum of a whole bank.
For illustrative purposes, we stick 
at the simplest case of the one-parameter family 
of (reduced) newtonian waveforms
in the range of chirp masses $(0.2M_\odot \leq {\cal M} \leq 10 M_\odot)$.
We assume the (initial) LIGO noise PSD \cite{LIGO_noise},
\beq
\Pi(f)=\frac{\Pi_0}{5}
\left\{
1+\left(\frac{f_0}{f}\right)^4+
2\left(\frac{f}{f_0}\right)^2
\right\},
\label{eq:ligonoise}
\eeq
with $f_0=300Hz$,  $f_i=40Hz$ and $f_s=400Hz$ 
and $\Pi_0$ is a constant of no concern to us here.
We sample the data at $3.2KHz$ ($4$-times the Nyquist rate),
and let $N_\Theta=2^{21}$, which corresponds
to the longest-lived source allowed (${\cal M}=0.2 M_{\odot}$).
Fig. 5 shows that the approximation provided by $\tilde{\Psi}_X^{(N)}$ is fairly good, 
and notably better than that obtainable from $\Psi_X^{(N)}$.
As $N$ (resp. $N_\Theta$) goes to infinity all approximants (\ref{eq:Rytov_like})
merge by construction into the Gumbel distribution.
This latter is indistinguishable from the independent-correlator approximation
in both Fig.s 4 and 5.
However, as seen from Fig.s 4 and 5, in most situations of practical interest 
for (possibly hierarchical) GW chirp search, eq.s (\ref{eq:Rytov_like})
would provide significantly better  approximants to the 
relevant CDF's than the plain independent-correlator (or Gumbel) approximation.
Being lower bounds, eq.s (\ref{eq:Rytov_like})   
are also conservative in terms of false-alarm probability, 
when used to set detection thresholds.\\
A sharper insight into the quality of the various
proposed approximants is gained by comparing these latter in terms
of the detection threshold corresponding to a given false-alarm
probability $\alpha$, as a function of $N_\Delta$.
This is illustrated in Fig. 6, for the special case $\alpha=10^{-3}$.
The thresholds in Fig. 6
correspond to assuming zero covariance among the correlators (dotted line, $\Phi_X^{(N)}$), 
and to including the covariance among nearest-neighbour correlators
along one (coalescency time, dashed line, $\Psi_X^{(N)}$)
or both (coalescency time and chirp mass, solid line, $\tilde{\Psi}_X^{(N)}$) 
coordinates of the newtonian parameter space, respectively.\\
In Fig. 6, we also include the limiting case 
where $\Gamma \approx \rho_M \approx 1$.
The corresponding (asymptotic) value of $\gamma$ (dotted-dashed line in Fig. 6) 
is obtained using the appropriate asymptotic form of $F_X(x)$,  
first quoted by Krolak  \cite{Jaranowski_et_Al} after
\cite{Adler} and \cite{Worsley}, viz.:
\beq
F_X(x) \sim
1-\pi^{-1}\mu[{\cal S}']
\left(\det[{\mathbf \Xi}]\right)^{1/2}
x \exp(-x),
\label{eq:asymp}
\eeq
where $\mu[\cdot]$ is the Lebesgue measure,
${\cal S}'$ is the allowed (reduced) source-parameter space,
and $\mathbf{\Xi}$ is the covariance
matrix of the derivative random-field \cite{explain}.

%%%%%%%%%%%%%%%%%%%%%%%%%%%%%%%%%%%%%%%%%
\section{The Minimal-Match Issue Revisited}
\label{sec:MM_again}
%%%%%%%%%%%%%%%%%%%%%%%%%%%%%%%%%%%%%%%%%

We are now in a position to re-address the question 
concerning the relationship
between the minimal match $\Gamma$ (and/or, the number of templates
in the parameter-space region of interest) and the undetected fraction
of potentially observable sources.
For simplicity, we shall again stick at the
simplest case of Newtonian waveforms and templates,
using the same range of source chirp masses and
antenna noise PSD as in Sect. \ref{subsec:Thresh}.

It should be noted first that a visible knee point  
in the $\Gamma$ vs. $N_\Delta$
curve invariably exists at some value $N_\Delta=N_*$,
beyond which {\em any} further increase in $N_\Delta$ yields only
a {\em modest} increase in $\Gamma$.\\
This is illustrated in Fig. 7, where
a knee point separating intervals 
of different (almost constant) slope
in the $\Gamma$ versus $N_\Delta$ curve
occurs at $N_\Delta \approx 5\cdot 10^3$.
Except for the numerical value of $N_*$, 
the existence of the above knee point 
is a {\em generic} feature of the $\Gamma$ vs. $N_{\Delta}$ curve, 
occurring also for more realistic, higher-order post-newtonian 
waveforms and templates.

Next we focus on eq. (\ref{eq:eta}),  
and use our {\em best} approximant 
$\tilde{\Psi}_X^{(N)}(x)$ in eq. (\ref{eq:Rytov_like})
for the no-signal CDF of the detection statistic
in order to compute the ratio 
$\gamma(\alpha,N_{max})/\gamma(\alpha,N)$  \cite{ratio}.
In Fig. 8  we accordingly compare $\eta_0$ (eq. (\ref{eq:eta0}), dashed line),
to $\eta$ (eq. (\ref{eq:eta}), solid lines), 
for $\alpha=10^{-k},~k=2,3,4$ (top-to bottom),
as functions of $N_\Delta$.
In Fig. 8 we assumed $\Gamma_{max}=0.99$ as a {\em bona fide}
practical limiting value, and computed $N_{max}$ accordingly.
From Fig. 8 it is seen that a knee point
in the curves exists at $N_\Delta \approx N_*\approx 5\cdot 10^3$. 
It is also seen that $\eta$ exceeds by a non negligible $5\%$ typical 
the plain estimate $\eta_0$ discussed in Sect. \ref{sec:MM_vs_undetect},
for $N_\Delta \approx N_*$.

It is also interesting to compare in terms of $\eta$
the alternative approximants  for the no-signal CDF 
of the detection statistic introduced and discussed in Sect. \ref{sec:Core}.
The resulting curves are shown in Fig. 9, where again 
we assumed $\Gamma_{max}=0.99$ as a {\em bona fide}
practical limiting value, and computed $N_{max}$ accordingly.
The curves  in Fig. 9
correspond to assuming zero covariance among the correlators (dotted line, $\Phi_X^{(N)}$), 
and to including the covariance among nearest-neighbour correlators
along one (coalescency time, dashed line, $\Psi_X^{(N)}$)
or both (coalescency time and chirp mass, solid line, $\tilde{\Psi}_X^{(N)}$) 
coordinates of the newtonian parameter space, respectively,
in  computing  $\gamma(\alpha,N_{max})$ in (\ref{eq:eta}).
For each case, the $\eta$ vs. $N_\Delta$ curves corresponding to
$\alpha=10^{-2},10^{-3},10^{-4}$ are displayed. 
It is seen that the dependence of $\eta$ on $\alpha$ becomes
the less relevant, the better approximant is used.

%%%%%%%%%%%%%%%%%%%%%%%%%%%%%%%%%%%%%%%%%
\section {Conclusions}
\label{sec:Conclu}
%%%%%%%%%%%%%%%%%%%%%%%%%%%%%%%%%%%%%%%%%

We presented a detailed analysis of the properties of the cumulative distribution
of the whole-bank supremum.
A number of tight lower-bound approximations, of increasing accuracy, have been introduced
and compared, valid for any (finite) number of templates. 
In addition, we studied the asymptotic properties as the number
of correlators goes to infinity, in the frame of extremal value theory (EVT).
We also shew how the
gaussian correlation inequality can be used to extend the validity
of simulations made on small-sized (toy) banks.

Concerning the minimal-match issue, 
we found a consistent evidence 
that increasing the number of templates beyond a critical value
does {\em not} produce a sensible increase in the detectable fraction
of potentially observable sources, at the expense of a marked growth
of computational load.
On the other hand, using the most accurate
representation of the no-signal cumulative distribution of the
(whole-bank) detection statistic to compute the threshold
results in a sizeable increase 
in the detectable fraction of potentially
observable sources over the naive $\propto \Gamma^3$ estimate.

%%%%%%%%%%%%%%%%%%%%%%%%%%%%%%%%%%%%%%%
\section*{Acknowledgements}
%%%%%%%%%%%%%%%%%%%%%%%%%%%%%%%%%%%%%%%
This work has been sponsored in part 
by the European Community through a Senior Visiting Scientist Grant 
to I.M. Pinto at NAO - Spacetime Astronomy Division, Tokyo, Japan, 
in connection with the TAMA project. 
I.M. Pinto wishes to thank all the TAMA staff at NAO, 
and in particular prof. Fujimoto Masa-Katsu 
and prof. Kawamura Seiji for kind hospitality 
and stimulating discussions.
%%%%%%%%%%%%%%%%%%%%%%%%%%%%%%%%%%%%%%%%

\newpage

\section*{Appendix A - Proof of Equation (\ref{eq:Bound_indep})}
%*************************************************************
%A - Independent case gives an lower bound for CDF
%*************************************************************
In this appendix we prove that the CDF of the whole-bank supremum 
for the independent-correlator case is a lower bound for the true one.
The proof relies on two preliminary Lemmas.
$$~$$
{\bf Lemma A.1} (Pitt's theorem \cite{Pitt})-
Let $\mu$ a centered gaussian measure 
\cite{Bogachev} in $\mathbf C$,
and let ${\cal A},{\cal B}$ two centered convex symmetric sets in $\mathbf C$. Then
\beq
\mu\left(
{\cal A} \cap {\cal B}
\right) \geq
\mu\left(
{\cal A}
\right) 
\mu\left(
{\cal B}
\right).
\eeq
$$~$$
{\bf Remark} - Pitt's theorem can be rephrased in functional (integral) form
as
\beq
E(f g) \geq E(f) E(g),
\label{eq:Pitt2}
\eeq
where $f,g$ are log-concave symmetric functions and $E(\cdot)$ is e.g.
the mean value w.r.t. to a centered complex gaussian measure $\mu$.
$$~$$
The following Lemma is a more or less obvious extension of the above
to ${\mathbf C}^N$.
$$~$$
{\bf Lemma A.2}  
Let $\mu_N$ a centered gaussian measure in ${\mathbf C}^N$. 
For any convex symmetric set ${\cal A}\in{\mathbf C}^N$, 
and any 
${\cal D}= \left\{
\vec{x} \in {\mathbf C}^N : |f(\vec{x})|\leq r 
\right\}$, 
where $f$ is any linear functional in ${\mathbf C}^N$ and $r \in {\mathbf
R}$,
one has
\beq
\mu_N\left({\cal A}\cap{\cal D}\right)
\geq
\mu_N({\cal A})\mu_N({\cal D}).
\label{eq:lem2}
\eeq
$$~$$     
{\bf Proof} - 
%A linear functional $f$ in ${\mathbf C}^N$
%can always be written as a hermitian scalar product, viz.:
%\beq
%f({\mathbf s})=<{\mathbf \lambda},{\mathbf s}>.
%\label{eq:isomorphi}
%\eeq 
%From (\ref{eq:isomorphi}) the set ${\cal D}$
%can be written, under a suitable unitary transformation
%(preserving obviously the gaussian character of the measure $\mu_N$):
It suffices to prove the Lemma assuming that $\mu_N$ is a {\em standard} gaussian
measure \cite{standard_gauss}.\\
Under a suitable unitary transformation
(preserving obviously the gaussian character of the measure $\mu_N$)
the set ${\cal D}$ can be transformed into: 
\beq 
{\cal D}=\left\{
\vec{y} \in {\mathbf C}^N : |y_1|\leq r
\right\}.
\eeq
Now let
\beq
\psi(z)=\int_{{\mathbf C}^{N-1}} I_{\cal A}(z,\vec{v})\mu_{N-1}(d\vec{v}),
\eeq
where $I_{\cal A}(z,\vec{v})$ denotes the indicator of ${\cal A}$, 
$z\in {\mathbf C}$ and $\vec{v}\in {\mathbf C}^{N-1}$.
$I_{\cal A}(z,\vec{v})$ is a non negative log-concave function, 
and $\mu_{N-1}$ is a non negative log-concave measure, 
whose related PDF is also non negative and log-concave.
Thus, according to theorem {\bf 4.6} in Bogachev \cite{Bogachev},
$\psi(z)$ is non negative and log-concave. 
The symmetry properties of ${\cal A}$ and $\mu_{N-1}$ imply 
that $\psi(z)$ is an even function.
The level-sets ${\psi(z)>r}$ are thus convex and symmetric.\\
Hence, letting ${\cal U}=\{z \in {\bf C}, |z|\leq r\}$, by eq. (\ref{eq:Pitt2}):
\beq
\int_{{\mathbf C}}\psi(z)I_{\cal U}(z)\mu_1(dz)
\geq
\int_{{\mathbf C}}\psi(z)\mu_1(dz)
\int_{{\mathbf C}}I_{\cal U}(z)\mu_1(dz)=
\mu_N({\cal A})\mu_N({\cal D}).
\label{eq:one}
\eeq 
On the other hand one has trivially:
\[
\int_{{\mathbf C}}\psi(z)I_{\cal U}(z)\mu_1(dz)
=
\int_{{\mathbf C}^N} I_{\cal A}(z,\vec{v}) I_{\cal U}(z)\mu_N(dz \times d\vec{v})
=\]
\beq
=\int_{{\mathbf C}^N}I_{{\cal A}\cap{\cal D}}(z,\vec{v})
\mu_N(dz \times d\vec{v})=\mu_N({\cal A}\cap{\cal D}).
\label{eq:two}
\eeq
Statement (\ref{eq:lem2}) follows from (\ref{eq:one}) and (\ref{eq:two})
$~\bullet$
$$~$$
We are now in a position to provide the
$$~$$
{\bf Proof of Eq. (\ref{eq:Bound_indep})} - 
%--insert \ref as appropriate ---%
Equation (\ref{eq:Bound_indep}) in Sect. \ref{sec:Core}
%--------------------------------%
can be rephrased as
\beq
\mbox{Prob}(|Y_1|\leq y_1,...,|Y_N|\leq y_N) \geq \prod_{i=1}^{N}\mbox{Prob}(|Y_i|\leq y_i),
\label{eq:ZZindbound}
\eeq
where $Y_i$,  $i=1,2,...N$ are centered jointly  
gaussian random variables in ${\mathbf C}^N$.
Let $\mu_N$ a centered  gaussian measure in ${\mathbf
C}^N$, and define the ({\em convex}, {\em symmetric}) sets:
\beq 
{\cal A}_i=
\left\{
\vec{a} \in {\mathbf C}^N : |a_i|\leq y_i \right\}.
\eeq
Inequality (\ref{eq:ZZindbound}) can be rewritten: 
\beq
\mu_N\left(\bigcap_{i=1}^{N} {\cal A}_i\right)
\geq
\prod_{i=1}^{N}\mu_N({\cal A}_i),
\label{eq:measure}
\eeq
which follows by induction from Lemma {\bf A.2}$~\bullet$
$$~$$
$$~$$
%*************************************************************
\section*{Appendix B - Asymptotic Properties of Bank Supremums CDF}
%*************************************************************
In this appendix we prove that the CDF of the supremum
of the bank correlator subsets corresponding to a fixed chirp mass,
as well as the CDF of the supremum of the whole bank
belong asymptotically to the Gumbel universal class, as
the pertinent number of correlators goes to infinity
(theorem 1 and 2, respectively).

%---------------------------------------------------
% Newtonian Approximation
%---------------------------------------------------

Throughout this Appendix we shall restrict
to the simplest  case of newtonian (0PN) waveforms
and templates, for which
\beq
\psi_T(f,\vec{\xi}_T)=
\frac{3}{128}
\left(
\frac{\pi G}{c^3}
\right)^{-5/3}
{\cal M}_T^{-5/3}
f^{-5/3}
\eeq
where ${\cal M}_T$ is the template chirp mass
\cite{chirp_mass} and $c$, $G$  have 
obvious meanings. 

It is convenient to introduce  the following
dimensionless variables and parameters:
\beq
\bar{f}=\frac{f}{f_i},~~~
\bar{\tau}=f_i \tau,~~~
\bar{\cal M}=\frac{\cal M}{M_{\odot}},~~~
\lambda=\frac{3}{128}
\left(\frac{\pi G f_i M_{\odot}}{c^3}\right)^{-5/3},
\label{eq:gammazzo}
\eeq
where $M_{\odot}$ is the solar mass.

The complex random field (\ref{eq:complex_Y})
and the pertinent {\em absolute} covariance (\ref{eq:quaqqua}) in turn become: 
\beq
Y(\tau_T,\xi_T)=Y(\bar{\tau},\bar{\cal M})
\label{eq:random_newt}
\eeq
and
\beq
|E[Y(\bar{\tau},\bar{\cal M})Y^*(\bar{\tau}',\bar{\cal M}')]|=
\frac{\displaystyle{
\left|
\int_{1}^{\bar{f}_s}
\bar{f}^{-7/3} \exp
\left\{j
\left[ 2\pi \bar{f} \Theta +
\lambda
\Delta
\bar{f}^{-5/3}
\right]
\right\}
\frac{d\bar{f}}{\Pi(\bar{f})}
\right|
}
}{
\displaystyle{
\left|
\int_{1}^{\bar{f}_s}
\bar{f}^{-7/3} 
\frac{d\bar{f}}{\Pi(\bar{f})}
\right|
}}
=:
\rho(\Theta,\Delta)
\label{eq:rhocaz}
\eeq
where $\Theta=\bar{\tau}'-\bar{\tau}$ and $\Delta=\bar{\cal M}'^{-5/3}-\bar{\cal M}^{-5/3}$.

Specializing eq. (\ref{eq:Y_DFT}) to the newtonian case, according to (\ref{eq:random_newt}) and (\ref{eq:rhocaz}), we introduce the 2D stationary sequence
\beq
Y_{h,k}=Y(h\delta_\Theta,k\delta_\Delta),~~~h,k \in {\mathbf N},
\label{eq:Y}
\eeq
and further define:
\beq
\rho_{m+1}=\rho(m\delta_{\Theta},0),
\eeq
\beq
\bar{\rho}_{m+1}=
\max_{h=1,\dots,N_{\theta}}
\rho(h\delta_{\Theta},m\delta_\Delta),
\eeq    
$\delta_\Theta$ and $\delta_\Delta$ being the discretization steps in $\bar{\tau}$
and $\bar{\cal M}$, respectively. 

We start by proving the following Lemma:
%Details will be published elsewhere.
$$~$$
%%%%%%%%%%%%%%%%%%%%%%%%%%%%%%%%%%%%%%%%%%%%%%%%%
% LEMMA B.1
%%%%%%%%%%%%%%%%%%%%%%%%%%%%%%%%%%%%%%%%%%%%%%%%%%
{\bf Lemma B.1} - If the equation
\beq
\Pi(\bar{f})+2\bar{f}\Pi'(\bar{f})=0,
\label{eq:eq_pierro}
\eeq
has at least one solution $\bar{f} \in [1,\bar{f}_s]$,
where $\Pi(\bar{f})$ is the antenna noise power spectral density, then
\beq
\bar{\rho}_{m}=o\left(\frac{1}{\sqrt{m}}\right). 
\label{eq:roots}
\eeq
Otherwise, 
\beq
\bar{\rho}_{m}=o\left(\frac{1}{m}\right).
\label{eq:no_roots}
\eeq
$$~$$
{\bf Proof} - Note first that:
\beq
\bar{\rho}_{m}\leq \max_{\Theta}\rho(\Theta,\Delta)=\rho[\Theta_{max}(\Delta),\Delta],
\label{eq:bnd_rght}
\eeq
where $\Theta$ on the r.h.s. is a continuous variable,
and $\Theta_{max}(\Delta)$ is the solution (in $\Theta$) of:
\beq
\frac{\partial \rho(\Theta,\Delta)}{\partial \Theta} = 0.
\label{eq:maxeq}
\eeq
The proof for the special case where (\ref{eq:eq_pierro}) has no root
in $[1,\bar{f}_s]$ follows immediately by Lebesgue theorem
(integration by parts).\\
The proof for the special case where (\ref{eq:eq_pierro}) has 
(at least) one root in $[1,\bar{f}_s]$ is readily obtained using 
the asymptotic stationary phase formula \cite{pha_staz}
to compute the correlator (\ref{eq:correlator}).
The phase stationary point in (\ref{eq:correlator}) is:
\beq
\bar{f}=\frac{5^{3/8}}{8\pi}
\left(
\frac{c^3}{G M_\odot}
\right)^{5/8}
\left(
\frac{\Delta}{\Theta}\right)^{3/8}=\bar{f}_a
\label{eq:fstaz}
\eeq
and the (absolute) correlation function can be accordingly written:
\beq
\rho(\Theta,\Delta)
\sim
4\pi^{4/3}
\sqrt{\frac{6}{5}}
\left(
\frac{GM_{\odot}}{c^3}
\right)^{5/6}
\Delta^{-1/2}
~
\frac{\bar{f}_a^{-1/2}}
{\Pi(\bar{f}_a)},
\label{eq:val_staz}
\eeq
where $\bar{f}_a$ depends on $\Theta$ (and $\Delta$) through eq.
(\ref{eq:fstaz}).
$~$
It is readily checked that $\partial \bar{f}_a/\partial\Theta$ is always
nonzero, so that the solution of (\ref{eq:maxeq}) are those of:
\beq
\frac{\partial \rho(\Theta,\Delta)}{\partial \bar{f}_a} = 0,
\eeq
which is nothing but eq. (\ref{eq:eq_pierro}), whose assumedly existing
solution will be denoted 
as $\bar{f}_a^\dag$.
Hence
\beq
\max_{\Theta}\rho(\Theta,\Delta)\sim
4\pi^{4/3}\sqrt{\frac{6}{5}}
\left(
\frac{GM_{\odot}}{c^3}
\right)^{5/6}
\Delta^{-1/2}                      
~
\frac
{\bar{f}_a^{\dag-1/2}}
{\Pi(\bar{f}_a^\dag)}.
\label{eq:asy_sup}
\eeq
The statement follows from (\ref{eq:asy_sup}) together with
(\ref{eq:bnd_rght}) $\bullet$
$$~$$
{\bf Remark} - For {\em all} first generation interferometers
equation (\ref{eq:eq_pierro}) has just one solution 
$\bar{f}_a^\dag \in [1,\bar{f}_s]$. Note that $\bar{f}_a^\dag$
does {\em not} depend on $\Delta$.
$$~$$
%%%%%%%%%%%%%%%%%%%%%%%%%%%%%%%%%%%%%%%%
%Lemma B.2
%%%%%%%%%%%%%%%%%%%%%%%%%%%%%%%%%%%%%%%%
{\bf Lemma B.2} - The following statement is true:
\beq
\rho_m=o\left(\frac{1}{m}\right).
\eeq
{\bf Proof} - The Lemma is based on the (stationary phase)
asymptotic evaluation for $\Theta \longrightarrow \infty$ of the integral:
\beq
\rho(\Theta,0)=
\left|
\frac{1}{{\cal N}}\int_{1}^{\bar{f}_s} \frac{\bar{f}^{-7/3}}{\Pi(\bar{f})}
\exp[j 2\pi  \bar{f} \Theta]d{\bar f}
\right|
=o\left(\frac{1}{\Theta}\right)~\bullet
\label{eq:stima}
\eeq
$$~$$
%*************************************************************
%CONDITION D in Castillo (chiamata 1)
%*************************************************************
We shall now introduce two ancillary conditions, which are suitable
weak (asymptotic) versions of the $M-$dependence  condition \cite{Castillo_conds}.
In particular condition \#1 is similar to the mixing condition
of dynamical systems \cite{Arnold}.
$$~$$
{\bf Condition 1} - Let $\{X_j\}$ a sequence of random variables, and let
$F_{j_1,j_2,...,j_r}(x_1,x_2,...,x_r)$ the joint CDF of $X_{j_1}, X_{j_2},...,X_{j_r}$. Condition \#1 holds if, 
for any set of integers 
$\{i_1<i_2,...<i_p\}$ and $\{j_1<j_2,...<j_q\}$ such that $j_1-i_p\ge s$,
and any real number $u$, a real-valued function $g(s)$ exists such that:
\beq
\left| F_{X_{i_1},...,X_{i_p},X_{j_1},...,X_{j_q}}(u,u,...,u)-
F_{X_{i_1},...,X_{i_p}}(u,u,...,u)F_{X_{j_1},...,X_{j_q}}(u,u,...,u) \right|
\le g(s),
\eeq
with 
\beq
\lim_{s\rightarrow \infty} g(s)=0.
\eeq
$$~$$
%*************************************************************
%CONDITION D' in Castillo (chiamata 2)
%*************************************************************
{\bf Condition 2:} Let $\{u_n\}$ a sequence of reals, and let $\{X_j\}$ a
stationary \cite{stat_what} random sequence. Condition  \#2 holds if:
\beq
\lim_{k \rightarrow \infty}\limsup_{n \rightarrow \infty} 
\left\{ n\sum_{j=2}^{ \lfloor n/k \rfloor}
{\mbox{Prob}\left[X_1>u_n, X_j>u_n \right]}\right\}=0.
\eeq
$$~$$
Conditions \#1 and \#2 above imply the following
$$~$$
%*************************************************************
%Enunciato Teorema Castillo 8.5
%*************************************************************
{\bf Lemma B.3} (\cite{Castillo}, theorem {\bf 8.5}) 
Let $\{X_j\}$ a stationary  sequence 
of random variables and let $\{a_n\}$, $\{b_n\}$ two sequences of reals. Assume that conditions \#1 and \#2 are satisfied with $u_n=a_n+b_nx$.
Let $\{X'_j\}$ a sequence of i.i.d. random variables having the same CDF as $\{X_j\}$, and ${\cal G}(x)$ any of the (three) universal extremal limit distributions \cite{Galambos}
of i.i.d. random variables. Then:
\beq
\lim_{n\rightarrow \infty} 
\mbox{Prob}\left[ \max_{j=1,...,n} X_j\le
a_n+b_n x\right]={\cal G}(x)
\label{eq:Cast1}
\eeq
and
\beq
\lim_{n\rightarrow \infty} 
\mbox{Prob}\left[ \max_{j=1,...,n} X'_j\le
a_n+b_n x\right]={\cal G}(x)
\label{eq:Cast2}
\eeq
are equivalent.
$$~$$
Lemma {\bf B.3} plays a key role. Loosely speaking,
it states that provided conditions \#1 and \#2 are satisfied, 
the supremum of a (stationary, {\em dependent}) sequence 
has the same CDF as if the sequence was made of independent r.v.\\
$$~$$
We are now in a position to prove three key Lemmas.
$$~$$
%*************************************************************
%Lemma B.4
%*************************************************************
{\bf Lemma B.4} - Let $G_i=Y_{i,k}$
with $k \in {\mathbf N}$ fixed,
and let $Z_i=|G_i|^2$. The $Z_i$ satisfy condition \#1.
$$~$$
{\bf Proof} - Let  $\{i_1<i_2,...<i_p\}$ and $\{j_1<j_2,...<j_q\}$ 
two set of integers such that $j_1-i_p \geq s$.
Let further ${\bf C}_p$ and ${\bf C}_q$ and ${\bf C}_{p+q}$ 
the correlation matrices of the zero-average complex gaussian random sequences
$\{G_{i_1},\dots,G_{i_p}\}$, $\{G_{j_1},\dots,G_{j_q}\}$, 
and $\{G_{i_1},\dots,G_{i_p},G_{j_1},\dots,G_{j_q}\}$, respectively.\\
Let ${\bf C}_1={\bf C}_{p+q}$ and ${\bf C}_2=$diag$[{\bf C}_p,{\bf C}_q]$,
and ${\bf G}=[G_{i_1},...,G_{i_p},G_{j_1},...,G_{j_q}]$, and let
\beq
F_{Z_{i_1},...,Z_{i_p},Z_{j_1},...,Z_{j_q}}(u,u,...,u)=
\frac{\omega}{\det^{p+q} {\bf C}_1}
\int_{\cal E}
e^{-{\bf G}^*{\bf C}^{-1}_1 {\bf G}}
d^{p+q}{\bf G}
\eeq
and
\beq
F_{Z_{i_1},...,Z_{i_p}}(u,u,...,u)F_{Z_{j_1},...,Z_{j_q}}(u,u,...,u)=
\frac{\omega}{\det^{p+q} {\bf C}_2}
\int_{\cal E}
e^{-{\bf G}^*{\bf C}^{-1}_2 {\bf G}}
d^{p+q}{\bf G}
\eeq
the pertinent cumulative distributions, 
$\omega$ being a suitable normalization constant, and 
${\cal E}$ the domain $ (|G_{i}| \le \sqrt{u} \cap |G_{j}| \le \sqrt{u} )$.\\
The $Z_i$ satisfy condition \#1 iff:
\beq
\lim_{s\rightarrow \infty}
\left| 
\frac{1}{\det^{p+q} {\bf C}_1}
\int_{\cal E}
e^{-{\bf G}^*{\bf C}^{-1}_1 {\bf G}}
d^{p+q}{\bf G}
-
\frac{1}{\det^{p+q}{\bf C}_2}
\int_{\cal E}
e^{-{\bf G}^*{\bf C}^{-1}_2 {\bf G}}
d^{p+q}{\bf G}
\right|=0.
\label{eq:condD}
\eeq
Under the assumption 
$$
\lim_{m\rightarrow\infty} E[G_i G^*_{i+m}]=
\lim_{m\rightarrow\infty} E[Y_{i,k}Y^*_{i+m,k}]=
\lim_{m\rightarrow\infty} \rho_{m+1}=0,~~~\forall i,
$$
which holds in view of Lemma {\bf B.2},
the off-diagonal blocks in ${\bf C}_1$ go to zero, and hence
\beq
\lim_{s\rightarrow \infty} ({\bf C}_1-{\bf C}_2)=0
~\Longrightarrow~
\lim_{s\rightarrow \infty} (\det{\bf C}_1-\det{\bf C}_2)=0.
\label{eq:hyp}
\eeq
To prove (\ref{eq:condD}) we note that:
\[
\left|
\frac{1}{\det^{p+q}{\bf C}_1} e^{-{\bf G^*}{\bf C}^{-1}_1{\bf G}}
-
\frac{1}{\det^{p+q}{\bf C}_2} e^{-{\bf G^*}{\bf C}^{-1}_2{\bf G}}
\right|
\le
\]
\[
=\frac{e^{-{\bf G^*}{\bf C}^{-1}_1{\bf G}}}{\det^{p+q} {\bf C}_1} 
\left[1-e^{-{\bf G^*}({\bf C}^{-1}_2-{\bf C}^{-1}_1){\bf G}}\right]
+
\frac{e^{-{\bf G^*}{\bf C}^{-1}_2{\bf G}}}{\det^{p+q} {\bf C}_1}
\left|
1-\left(\frac{\det {\bf C}_1}{\det {\bf C}_2}\right)^{p+q}
\right|\le
\]
\beq
\le \frac{e^{-{\bf G^*}{\bf C}^{-1}_1{\bf G}}}{\det^{p+q} {\bf C}_1}
\left[
e^{|{\bf G}|^2 ||{\bf C}^{-1}_2-{\bf C}^{-1}_1 ||}-1
\right]
+
\frac{e^{-{\bf G^*}{\bf C}^{-1}_2{\bf G}}}{\det^{p+q} {\bf C}_1}
\left|
1-\left(\frac{\det {\bf C}_1}{\det {\bf C}_2}\right)^{p+q}
\right|,
\label{eq:interme}
\eeq
where $||\cdot||$ is the (matrix) $L_2$ norm.
Using further the simple inequality
 $|{\bf G}|^2\le\sum_{h=1}^p|G_{i_h}|^2+
\sum_{k=1}^q|G_{j_k}|^2\le(p+q)u$, 
we obtain from (\ref{eq:interme}):
\[
\left|
\frac{1}{\det^{p+q}{\bf C}_1} e^{-{\bf G^*}{\bf C}^{-1}_1{\bf G}}
-
\frac{1}{\det^{p+q}{\bf C}_2} e^{-{\bf G^*}{\bf C}^{-1}_2{\bf G}}
\right|
\le
\]
\beq
\le
\frac{1}{\det^{p+q} {\bf C}_1}
\left\{
\left[
e^{(p+q)u ||{\bf C}^{-1}_2-{\bf C}^{-1}_1 ||}-1
\right]
+
\left|
1-\left(\frac{\det {\bf C}_1}{\det {\bf C}_2}\right)^{p+q}
\right|
\right\}
\Longrightarrow_{s\rightarrow\infty} 0,
\label{eq:final}
\eeq
in view of (\ref{eq:hyp}).
Equation (\ref{eq:condD}) follows $\bullet$
$$~$$
The next Lemma {\bf B.5} is needed to prove Lemma {\bf B.6}.
$$~$$
%*************************************************************
%Lemma B.5
%*************************************************************
{\bf Lemma B.5} - Let $G_1$ and $G_2$ two complex standard gaussian random variables, and let:
\beq
c=\left|E\left[G_1 G^*_2\right]\right|<1.
\eeq
The following inequality holds:
\beq
\mbox{Prob}\left[|G_1|^2>u, |G_2|^2>u\right]\le
\frac{1+c}{1-c} e^{-\frac{2}{1+c}u}.
\eeq
{\bf Proof}  - From the standard definition of bivariate 
(complex) normal variables,
\beq
\mbox{Prob}\left[|G_1|^2>u, |G_2|^2>u\right]=
\frac{4}{1-c^2}   
\int_{\sqrt u}^{\infty}\int_{\sqrt u}^{\infty}
x_1 x_2 e^{-\frac{x_1^2+x_2^2}{1-c^2}}
I_0\left(\frac{2c}{1-c^2}x_1 x_2\right)dx_1dx_2,
\eeq
where $I_0(\cdot)$ is the modified Bessel function. 
Representing this latter in integral form \cite{AS}, gives:
\beq
\mbox{Prob}\left[|G_1|^2>u, |G_2|^2>u\right]\le
\frac{4}{1-c^2}
\int_{\sqrt u}^{\infty}\int_{\sqrt u}^{\infty}
x_1 x_2 e^{-\frac{x_1^2+x_2^2-2c x_1 x_2}{1-c^2}}
dx_1dx_2.
\eeq
Switching to polar coordinates ($\rho$, $\phi$),
\beq
x_1^2+x_2^2-2c x_1 x_2=\rho^2(1-c\sin 2\phi)\ge\rho^2(1-c),
\eeq
whence:
\[
\mbox{Prob}\left[|G_1|^2>u, |G_2|^2>u\right]\le
\frac{4}{1-c^2}
\int_{\sqrt u}^{\infty}\int_{\sqrt u}^{\infty}
x_1 x_2 e^{-\frac{x_1^2+x_2^2}{1-c^2}(1-c)}
dx_1dx_2=
\]
\beq
=\frac{4}{1-c^2}
\int_{\sqrt u}^{\infty}\int_{\sqrt u}^{\infty}
x_1 x_2 e^{-\frac{x_1^2+x_2^2}{1+c}}
dx_1dx_2=
\frac{4}{1-c^2}
\left[\int_{\sqrt u}^{\infty}
x_1 e^{-\frac{x_1^2}{1+c}}
dx_1\right]^2
=\frac{1+c}{1-c}e^{-2\frac{u}{1+c}},
\label{eq:cazz}
\eeq
q.e.d. $\bullet$
$$~$$
%*************************************************************
%Lemma B.6
%*************************************************************
{\bf Lemma B.6} - Let  $\{H_j\}$ a sequence 
of complex standard gaussian random variables, 
let $u_n=\log n+x$, and let:
\beq
\gamma_{m+1}=\left| E\left[ H_1
H^*_{1+m} \right] \right|<1 ,~~ m\neq 0.
\label{eq:gamma}
\eeq 
If, for some $0<\beta<\infty$
\beq
\gamma_{m}=o\left(\frac{1}{m^{\beta}}\right),
\label{eq:gammalimite}
\eeq
then:
\beq
\lim_{k \rightarrow \infty}\limsup_{n \rightarrow \infty} 
\left\{n
\sum_{j=2}^{\lfloor n/k \rfloor}{\mbox{Prob}\left[\left|H_1\right|^2>u_n, \left|H_j\right|^2>u_n \right]}\right\}=0.
\label{eq:tesi}
\eeq
$$~$$
{\bf Proof} - In view of Lemma {\bf B.5},
\beq
S_n=n \sum_{j=2}^{\lfloor n/k \rfloor}{\mbox{Prob}\left[\left|H_1\right|^2>\log n +x ,
\left|H_j\right|^2>\log n +x \right]}\le
n M \sum_{j=2}^{\lfloor n/k \rfloor}e^{-\frac{2}{\gamma_j+1}\log n}
=
M \sum_{j=2}^{\lfloor n/k \rfloor}n^{\frac{\gamma_j-1}{\gamma_j+1}},
\label{eq:somma}
\eeq                                             
where:
\beq
M=\frac{1+\bar \gamma}{1-\bar \gamma} e^{2|x|}, ~~~ \bar \gamma=\sup_{j\ge 2}\gamma_j<1.
\label{eq:defs}
\eeq
The last inequality follows from $\gamma_j<1$~~~$\forall j\ge 2$, 
and from the asymptotic decay rate of $\gamma_j$.
\\
From eq.(\ref{eq:somma}) one readily gets:
\beq
S_n\le M\sum_{j=2}^{\lfloor n/k \rfloor}{\lfloor n/k \rfloor}^{\frac{\gamma_j-1}{\gamma_j+1}}k^{\frac{\gamma_j-1}{\gamma_j+1}}.
\label{eq:minorati}
\eeq
Letting in (\ref{eq:minorati}) $m=\lfloor n/k \rfloor$, and using the obvious inequality:
\beq
k^{\frac{\gamma_j-1}{\gamma_j+1}}\le
k^{\frac{\bar\gamma-1}{\bar\gamma+1}}
~~~\forall j\ge 2,
\eeq
eq. (\ref{eq:somma}) becomes:
\beq
S_n\le M
k^{\frac{\bar\gamma-1}{\bar\gamma+1}}
\sum_{j=2}^m m^{\frac{\gamma_j-1}{\gamma_j+1}}.
\eeq
The Lemma now readily follows from the fact that:
\beq
\lim_{m\rightarrow\infty}
\sum_{j=2}^m m^{\frac{\gamma_j-1}{\gamma_j+1}}=1.
\eeq
Indeed letting:
\beq
\bar S_m=\sum_{j=2}^m m^{\frac{\gamma_j-1}{\gamma_j+1}}-1
=\sum_{j=2}^m \left(
m^{\frac{\gamma_j-1}{\gamma_j+1}}-\frac{1}{m}
\right)-\frac{1}{m},
\eeq
and taking limit $m\rightarrow\infty$, one gets:
\beq     
\bar S_m\sim
\sum_{j=2}^m \left(
m^{\frac{\gamma_j-1}{\gamma_j+1}}-\frac{1}{m}
\right)=
\sum_{j=2}^m \frac{
e^{\frac{2\gamma_j}{\gamma_j+1}\log m}-1}{m}.
\eeq
Obviously,
\beq
\forall \delta>1,~~~ \exists \zeta_0: e^\zeta-1<\delta\zeta,
~~~\forall \zeta\in [0,\zeta_0].
\eeq
Now let $j'$ the {\em smallest} integer such that: 
\beq
\frac{2\gamma_j}{1+\gamma_j}
\log m\le\zeta_0~~~\forall j\ge j'.
\label{eq:jj'}
\eeq
Then:
\beq
\frac{2\gamma_{j'-1}}{\gamma_{j'-1}+1}\log m > \zeta_0,
\label{eq:prima}
\eeq
and
\beq
e^{\frac{2\gamma_j}{\gamma_j+1}\log m}-1<
\frac{2\delta \gamma_j}{1+\gamma_j}\log m,
~~~\forall j\ge j'.
\label{eq:seconda}
\eeq
Let:
\beq
\bar S_m
\sim 
\sum_{j=2}^{j'-1} \frac{
e^{\frac{2\gamma_j}{\gamma_j+1}\log m}-1}
{m}
+
\sum_{j=j'}^{m}
\frac{
e^{\frac{2\gamma_j}{\gamma_j+1}\log m}-1}{m}.
\label{eq:trikini}
\eeq
We can use the inequality (\ref{eq:seconda})
to show that the second partial sum in (\ref{eq:trikini})
vanishes in the limit $m \rightarrow \infty$.\\
Indeed:
\[
\sum_{j=j'}^{m}
\frac{
e^{\frac{2\gamma_j}{\gamma_j+1}\log m}-1}
{m}\leq
\frac{2\delta\log m}{m}\sum_{j=j'}^{m}\frac{\gamma_j}{1+\gamma_j}
\le
\frac{2\delta\log m}{m}\max_{j=j',...,m}
\left(\gamma_j j^{\beta}\right)\sum_{j=j'}^{m}
j^{-\beta}
\]
\beq
\le
\frac{2\delta\log m}{m}\max_{j=j',...,m}
\left(\gamma_j j^{\beta}\right)
\int_1^m  j^{-\beta} dj=
\frac{2\delta\log m}{m}\max_{j=j',...,m}
\left(\gamma_j j^{\beta}\right)
\left(\frac{m^{1-\beta}-1}{1-\beta}\right)
\label{eq:bikini}.
\eeq
On the other hand in view of (\ref{eq:defs}):
\[
\sum_{j=2}^{j'-1} \frac{
e^{\frac{2\gamma_j}{\gamma_j+1}\log m}-1}
{m}
\le
(j'-2)\frac{e^{\frac{2\bar\gamma}{1+\bar\gamma}\log m}-1}{m}
=\frac{(j'-2)(\gamma_{j'-1})^{1/\beta}}{(\gamma_{j'-1})^{1/\beta}}
\left(m^{\frac{\bar\gamma-1}{\bar\gamma+1}}-\frac{1}{m}\right).
\]
Now we can use the inequality (\ref{eq:prima}) to show that the first partial sum in
(\ref{eq:trikini}) also vanishes in the limit $m \rightarrow \infty$:
\beq
\sum_{j=2}^{j'-1} \frac{
e^{\frac{2\gamma_j}{\gamma_j+1}\log m}-1}
{m}
\le
(j'-2)(\gamma_{j'-1})^{1/\beta}
\left(\frac{2\log m}{\zeta_0}-1\right)^{1/\beta}
\left(m^{\frac{\bar\gamma-1}{\bar\gamma+1}}-\frac{1}{m}\right).
\label{eq:primoterm}
\eeq
This completes the proof $\bullet$
$$~$$
{\bf Remark} - Lemma {\bf B.6} holds for general (i.e., stationary 
as well as non-stationary) sequences. This most general case will be
invoked in Lemma {\bf B.8}.
$$~$$
From Lemmas {\bf B.3} to {\bf B.6}, one readily proves that
the CDF of the supremum of the whole-bank subsets corresponding to a fixed chirp mass 
belongs asymptotically to the Gumbel universal class, as
the pertinent number of correlators $N_{\Theta}$ goes to infinity \cite{remark}.
$$~$$
%*************************************************************
%Teorema 1
%*************************************************************
{\bf Theorem 1}  - Let $Z_j=|Y_{j,k}|^2$, $k \in {\mathbf N}$ being fixed.
If, for some $0<\beta<\infty$
\beq
\rho_{m}=o\left(\frac{1}{m^\beta}\right),
\label{eq:hyppo1}
\eeq
then the supremum of $\{Z_j|j=1,\dots,N_{\Theta}\}$ is asymptotically
Gumbel-distributed for $N_{\Theta} \rightarrow \infty$, viz.:
\beq
\lim_{N_{\Theta} \rightarrow \infty}
\mbox{Prob}\left[ \max_{j=1,...,N_{\Theta}} Z_j \le
\log N_{\Theta}+x\right]=\exp[-\exp(-x)],
~~~\forall k.
\label{eq:th1}
\eeq
$$~$$
{\bf Proof} - 
The $Z_j$ are exponentially distributed.
If they were independent, eq. (\ref{eq:th1}) would be trivial,
as noted in Sect. \ref{sec:Core}.
However, $\forall k \in {\mathbf N}$, the sequence
$Z_j$ satisfies both conditions \#1 and \#2, with $u_n=\log n+x$,
in view of Lemmas {\bf B.4} and {\bf B.6}.
Thus by Lemma {\bf B.3},  eq. (\ref{eq:th1}) holds true despite
the $Z_j$ being dependent $\bullet$
$$~$$
We turn now to prove Theorem {\bf 2}. To this end we start
with two relevant Lemmas.
$$~$$
%*************************************************************
%Lemma B.7
%*************************************************************
{\bf Lemma B.7} - Let:
\beq	
\displaystyle{
{\cal Z}_k=
\max_{j=1,\dots,N_{\Theta}} 
|Y_{j,k}|^2}.
\label{eq:cal_zeta}
\eeq
The sequence $\{{\cal Z}_k\}$ satisfies condition \#1.
$$~$$
{\bf Proof} - From the obvious relation 
\beq
\mbox{Prob}\left[{\cal Z}_k\le u\right]=
\mbox{Prob}\left[|Y_{1,k}|^2\le u,|Y_{2,k}|^2\le u,..., |Y_{N_{\Theta},k}|^2\le u\right],
\eeq
the probability $\mbox{Prob}\left[{\cal Z}_k\le u\right]$ is a joint probability of gaussian 
complex variables. The proof follows from Lemma {\bf B.4} $\bullet$
$$~$$
%*************************************************************
%Lemma B.8
%*************************************************************
{\bf Lemma B.8} - If, for some $0<\bar\beta<\infty$
\beq
\bar{\rho}_{m}=o\left(\frac{1}{m^{\bar\beta}}\right),
\label{eq:gammalimite1}
\eeq
then the sequence  $\{{\cal Z}_k\}$ 
satisfies condition \#2, with $u_n=\log n+x$.
$$~$$
{\bf Proof} - Let 
\beq
E_h=
\left\{
\begin{array}{l}
\mbox{True}, {\cal Z}_h > u_n \\
\mbox{False}, {\cal Z}_h \leq u_n
\end{array},
\right.
~h=1,...,N_\Delta,
\eeq
and let 
\beq
E_{j,h}=
\left\{
\begin{array}{l}
\mbox{True}, |Y_{j,h}|^2>u_n \\
\mbox{False},|Y_{j,h}|^2\leq u_n
\end{array},
\right.
~j=1,...,N_\Theta ~h=1,...,N_\Delta.
\eeq
Clearly $E_h=\displaystyle{\bigcup_{j=1}^{N_{\Theta}}} E_{j,h}$, and hence:
\beq
\mbox{Prob}[E_1\cap E_k]=
\mbox{Prob}\left[
\left(\bigcup_{i=1}^{N_{\Theta}} E_{i,1}\right)
\cap
\left(\bigcup_{j=1}^{N_{\Theta}} E_{j,k}\right)
\right]=
\mbox{Prob}\left[\bigcup_{i,j=1}^{N_{\Theta}}
\left(E_{i,1}\cap E_{j,k}
\right)
\right]\le
\sum_{i,j=1}^{N_{\Theta}}
\mbox{Prob}\left[E_{i,1}\cap E_{j,k}\right].
\eeq
Now condition \#2  for ${\cal Z}_k$ can be written:
\beq
\lim_{m \rightarrow \infty}\limsup_{n \rightarrow \infty}\left\{ n
\sum_{k=2}^{\lfloor n/m \rfloor}
{\mbox{Prob}\left[{\cal Z}_1>u_n, {\cal Z}_k>u_n \right]}\right\}=0.
\eeq
On the other hand,
\beq
\lim_{m \rightarrow \infty}\limsup_{n \rightarrow \infty} 
\left\{n
\sum_{k=2}^{\lfloor n/m \rfloor}
{\mbox{Prob}\left[{\cal Z}_1>u_n, {\cal Z}_k>u_n \right]}
\right\}
\le
\sum_{i,j=1}^{N_{\Theta}}
\lim_{m \rightarrow \infty}\limsup_{n \rightarrow \infty} 
\left\{n \sum_{k=2}^{\lfloor n/m \rfloor}
{\mbox{Prob}\left[E_{i,1}\cap 
E_{j,k} \right]}
\right\}.
\eeq
But in view of Lemma {\bf B.6}, 
\beq
\lim_{m \rightarrow \infty}\limsup_{n \rightarrow \infty} \left\{n
\sum_{k=2}^{\lfloor n/m \rfloor}
{\mbox{Prob}\left[E_{i,1}\cap 
E_{j,k} \right]}\right\}=0,~~~\forall i,j.
\label{eq:finalmente}
\eeq
and the proof is complete $\bullet$
$$~$$
From Lemmas {\bf B7} and {\bf B8}, we can finally prove
$$~$$
%*************************************************************
%Teorema 2
%*************************************************************
{\bf Theorem 2} - If, for some $0<\bar\beta<\infty$
\beq
\bar{\rho}_{m}=o\left(\frac{1}{m^{\bar\beta}}\right),
\eeq
then the supremum of $\{{\cal Z}_k|k=1,\dots,N_{\Delta}\}$ is asymptotically Gumbel distributed for $N_{\Delta}\rightarrow\infty$, viz.:
\beq
\lim_{N_{\Delta}\rightarrow\infty}
\mbox{Prob}
\left[
\max_{k=1,\dots,N_{\Delta}}
{\cal Z}_k=
\max_{k=1,\dots,N_{\Delta}}
\max_{j=1,\dots,N_{\Theta}}
|Y_{j,k}|^2
\leq
\log (N_{\Theta} N_{\Delta}) + x
\right]=\exp[-\exp(-x)].
\eeq
$$~$$
{\bf Proof} - The sequence $\{{\cal Z}_k\}$ satisfies both
conditions \#1 and \#2, and thus by Lemma {\bf B.3} 
\beq
\lim_{N_{\Delta}\rightarrow\infty}
\mbox{Prob}\left[\max_{k=1,...,N_{\Delta}}{\cal Z}_k
\le
\log (N_{\Theta}N_{\Delta})+x\right]
=
\lim_{N_{\Delta}\rightarrow\infty}
\left[F_{{\cal Z}_k}\left(
\log (N_{\Theta}N_{\Delta})+x
\right)\right]^{N_{\Delta}}.
\label{eq:fine}
\eeq
The explicit form of the r.h.s. of (\ref{eq:fine}), which corresponds to
the case where the ${\cal Z}_k$ are assumed as independent is as
yet unknown, such being the CDF $F_{{\cal Z}_k}$.
However, by  the inclusion-exclusion principle \cite{Shirayev}:
\beq
1-F_{{\cal Z}_k}(x)=
N_{\Theta}
\mbox{Prob}\left[Z_{j}>x\right]
-\sum_{i>j} 
\mbox{Prob}\left[Z_i>x,Z_j>x\right]+
\sum_{i>j>h} 
\mbox{Prob}\left[Z_i>x,Z_j>x,Z_h>x\right]-...
\label{eq:unomeneffe}
\eeq
In the limit as $x \rightarrow \infty$, the r.h.s. of (\ref{eq:unomeneffe})
is infinitesimal. The first term in (\ref{eq:unomeneffe}) is the principal part.
Indeed, for any (non empty) set of integers  
$\left\{i_1,i_2,...,i_p\right\}$ different from $i$ and $j$, and for all $x$
\beq
\mbox{Prob}\left[Z_i>x,Z_j>x\right]\ge
\mbox{Prob}\left[Z_i>x,Z_j>x,Z_{i_1}>x,...,Z_{i_p}>x\right].
\eeq
Furthermore, from Lemma {\bf B.5},
\beq
\mbox{Prob}\left[Z_i>x,Z_j>x\right] \sim \exp(-w x),~~~w > 1,
\eeq
while 
\beq
\mbox{Prob}\left[Z_{j}>x\right] = \exp(-x).
\eeq
Hence, 
\beq
\lim_{x\rightarrow\infty}
\frac
{1-F_{{\cal Z}_k}(x)}
{1-\exp[-N_{\Theta}\exp(-x)]}
=
\lim_{x\rightarrow\infty}
\frac
{N_{\Theta}\exp(-x)}
{N_{\Theta}\exp(-x)}
=1,
\label{eq:righttail}
\eeq
$$~$$
which means that the two distributions
$F_{{\cal Z}_k}(x)$ and $\exp[-N_{\Theta}\exp(-x)]$
are {\em right tail equivalent}.
Under this assumption, theorem {\bf 3.15} in \cite{Castillo}
states that the domain of attraction of the two 
distributions is the same, viz:

\beq
\lim_{N_{\Delta}\rightarrow\infty}
\left\{F_{{\cal Z}_k}\left[
\log (N_{\Theta} N_{\Delta})+x
\right]\right\}^{N_{\Delta}}
=
\lim_{N_{\Delta}\rightarrow\infty}
\left\{
\exp
\left[-N_{\Theta}\exp
\left(-\log(N_{\Theta}N_{\Delta})-x
\right)
\right]
\right\}^{N_{\Delta}}=\exp[-\exp(-x)].
\label{eq:asymcaz}
\eeq
Letting (\ref{eq:asymcaz}) into (\ref{eq:fine}) proves the theorem $\bullet$
$$~$$
{\bf Remark} - Theorem {\bf 2} shows that (uniform) convergence is achieved 
in the limit $N_{\Delta}\rightarrow\infty$ at any fixed $N_{\Theta}$.
We add without proof that the larger $N_{\Theta}$, the faster   
the convergency rate.
  
\newpage

%%%%%%%%%%%%%%%%%%%%%%%%%%%%%%%%%%%%%%%%%%%%%%%%%%%%%%%%%
\begin{center}
{\bf Captions to the Figures}
\end{center}
%%%%%%%%%%%%%%%%%%%%%%%%%%%%%%%%%%%%%%%%%%%%%%%%%%%%%%%%%
$$
~
$$
Fig. 1 - Newtonian template bank with lexicographic ordering.
$$
~
$$
Fig. 2 - Difference between the Gumbel distribution $G^{N}(x)$ and the cumulative distribution 
of the whole-bank supremum $\Phi_X^{(N)}(x)$ corresponding to the assumption 
of statistically independent correlators vs. $x-\log N$ for various values of $N$. 
The supremum of the difference is plotted in the inset vs. $\log_2 N$. 
$$
~
$$
Fig. 3 - Difference between the Gumbel distribution $G^{N_\Theta}(x)$ and the
cumulative distribution of a single reduced correlator vs. $x-\log N_\Theta$, 
for several values of $N_{\Theta}$, and two different values of $\rho_M$.
$$
~
$$
Fig. 4 - Comparison among several approximants for the cumulative distribution 
of a single reduced correlator vs. $x-\log N_\Theta$.
Numerical simulation ($10^4$ realizations).
$$
~
$$
Fig. 5 - Comparison among several approximants for the cumulative distribution 
of the whole-bank supremum vs. $x-\log N_\Theta$.
Numerical simulation ($10^4$ realizations).
$$
~
$$
Fig. 6 - Detection threshold $\gamma$ corresponding to $\alpha=10^{-3}$
obtained from different models of the no-signal cumulative distribution 
of the whole-bank supremum vs. number of reduced templates $N_\Delta$.
The dotted-dashed line is the continuous limit, eq. (\ref{eq:asymp}).
Search range $0.2{\cal M}_{\odot} \leq {\cal M}  \leq 10{\cal M}_{\odot}$,
Newtonian waveforms, Ligo-I noise.
$$
~
$$
Fig. 7 -   Minimal match $\Gamma$ vs. number of reduced templates $N_\Delta$. 
Search range $0.2{\cal M}_{\odot} \leq {\cal M}  \leq 10{\cal M}_{\odot}$,
Newtonian waveforms, Ligo-I noise.
$$
~
$$
Fig. 8 - Detectable fractions $\eta$ (eq. (\ref{eq:eta}), solid line) 
and $\eta_0$ (eq. (\ref{eq:eta0}), dashed line) of potentially observable sources 
vs. number of reduced templates $N_\Delta$, for several false-alarm levels.
Search range $0.2{\cal M}_{\odot} \leq {\cal M}  \leq 10{\cal M}_{\odot}$,
Newtonian waveforms, Ligo-I noise.
$$
~
$$
Fig. 9 - Detectable fractions $\eta$ of potentially observable sources, 
obtained from different models of the no-signal cumulative distribution 
of the whole-bank supremum
vs. number of reduced templates $N_\Delta$, for several false-alarm levels. 
Search range $0.2{\cal M}_{\odot} \leq {\cal M}  \leq 10{\cal M}_{\odot}$,
Newtonian waveforms, Ligo-I noise.


\begin{thebibliography}{99}
\bibitem{DFT}{The $DFT$ definition is here intended to embody the sampling step, 
so as to make $DFT_n[\cdot]\sim{\cal F}_{x\rightarrow f_n}[\cdot]$. }
\bibitem{corr_neigh}{In \cite{Jaranowski_et_Al} the boundary of  the correlation 
neighbourhood of a template is more or less arbitrarily defined as the locus where 
the overlap equals $1/2$.}
\bibitem{hierarchical}{A possible exception to this may occur in multi-step hierarchical searches.}
\bibitem{heuristic}{We note that when
the covariance among the correlators is zero,
$$
\mbox{Prob}[{\cal Z}_{j+1}>x|{\cal Z}_{j}>x, \mbox{no signal}]=
\mbox{Prob}[{\cal Z}_{j+1}>x| \mbox{no signal}] \geq
$$
$$
\geq
\mbox{Prob}[Z_{i,j+1}>x| \mbox{no signal}]=
\mbox{Prob}[Z_{i,j+1}>x|Z_{m,j}>x, \mbox{no signal}],
$$
and hence  (\ref{eq:ansatz}) is verified. 
On the other hand, in the opposite limiting case 
where the covariance among the correlators is one,
both sides of (\ref{eq:ansatz}) are equal to one
and (\ref{eq:ansatz}) still holds.}
\bibitem{ratio}{Note that $\gamma(\alpha,N_{max})/\gamma(\alpha,N) \geq 1$,
since $\gamma(\alpha,N)$ is a non-decreasing function of $N$,
whatever $\alpha$.}
\bibitem{chirp_mass}{The chirp-mass is ${\cal M}=\mu^{3/5}M_{tot}^{2/5}$,
where $\mu$ and $M_{tot}$ are the reduced and total mass of the source, respectively.}
\bibitem{pha_staz}{The relevant large asymptotic parameter is the ratio $cT_o/\pi r_g$,  where $T_0$ and $r_g$ are the source (initial) orbital period and gravitational radius.}
\bibitem{Castillo_conds}{These are related to the weak mixing conditions
$D$ and $D'$ quoted in  Castillo's book \cite{Castillo}.}
\bibitem{stat_what}{The required stationarity here refers to the joint CDFs,
and is implied for gaussian correlators by the stationarity 
of the correlation matrix .}
\bibitem{remark}{One could readily prove that also the supremum of the
whole-bank subsets corresponding to a fixed coalescency time 
belongs asymptotically to the Gumbel universal class, as
the pertinent number of correlators $N_{\Delta}$ goes to infinity.
To this end, it is sufficient to rephrase Lemma {\bf B.2} by noting
that
$$
\tilde{\rho}_m=o\left(\frac{1}{m}\right),
$$
where
\beq
\tilde{\rho}_{m+1}=|E[Y_{j,k}Y^*_{j,k+m}]|,
\eeq
which follows from the asymptotic (stationary phase)
evaluation for $\Delta \longrightarrow \infty$ 
of the integral :
$$
\rho(0,\Delta)=
\left|
\frac{1}{{\cal N}}\int_{1}^{\bar{f}_s} \frac{\bar{f}^{-7/3}}{\Pi(\bar{f})}
\exp[j( \lambda \bar{f}^{-5/3}\Delta)]d\bar{f}
\right|
=o\left(\frac{1}{\Delta}\right).
$$
Rephrasing the subsequent Lemmas {\bf B.3} to {\bf B.6} as well as Theorem 1 is straightforward.}
\end{thebibliography}
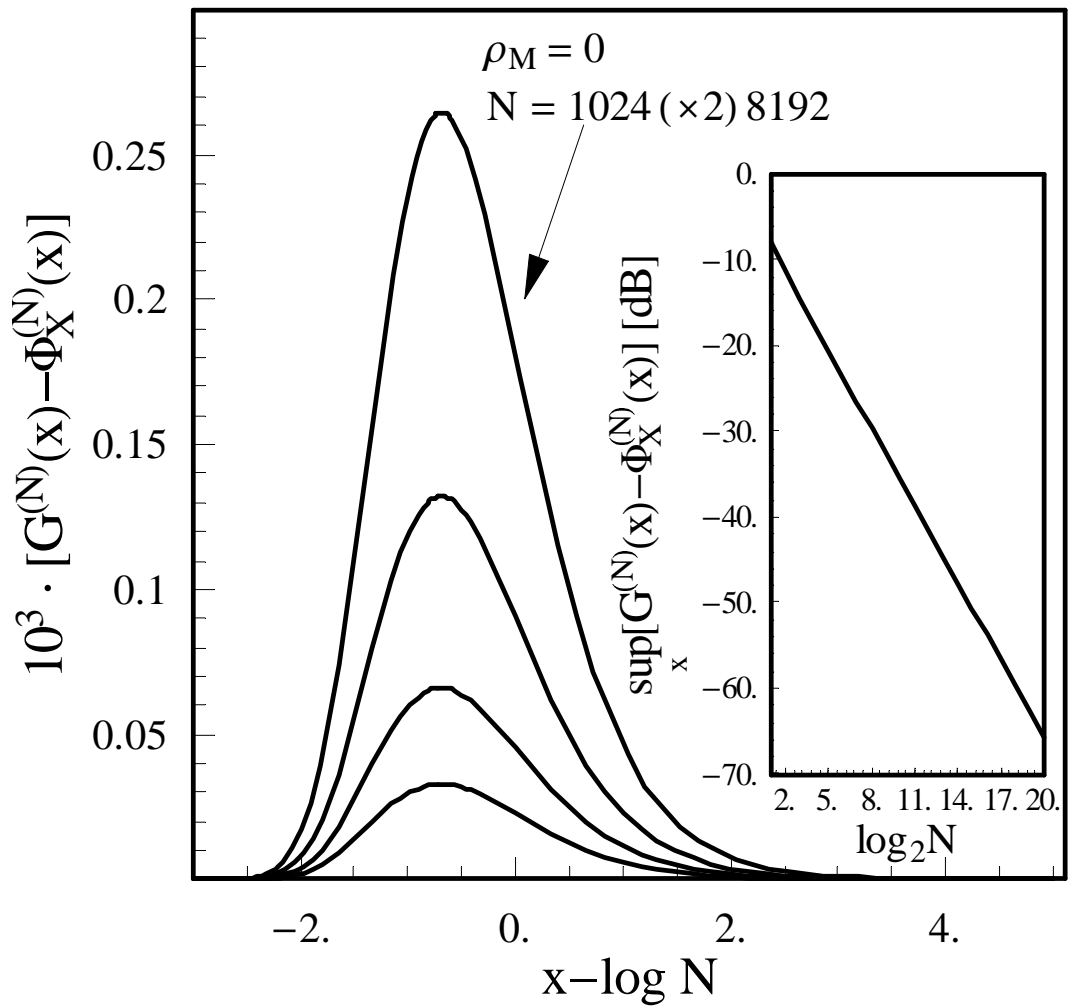
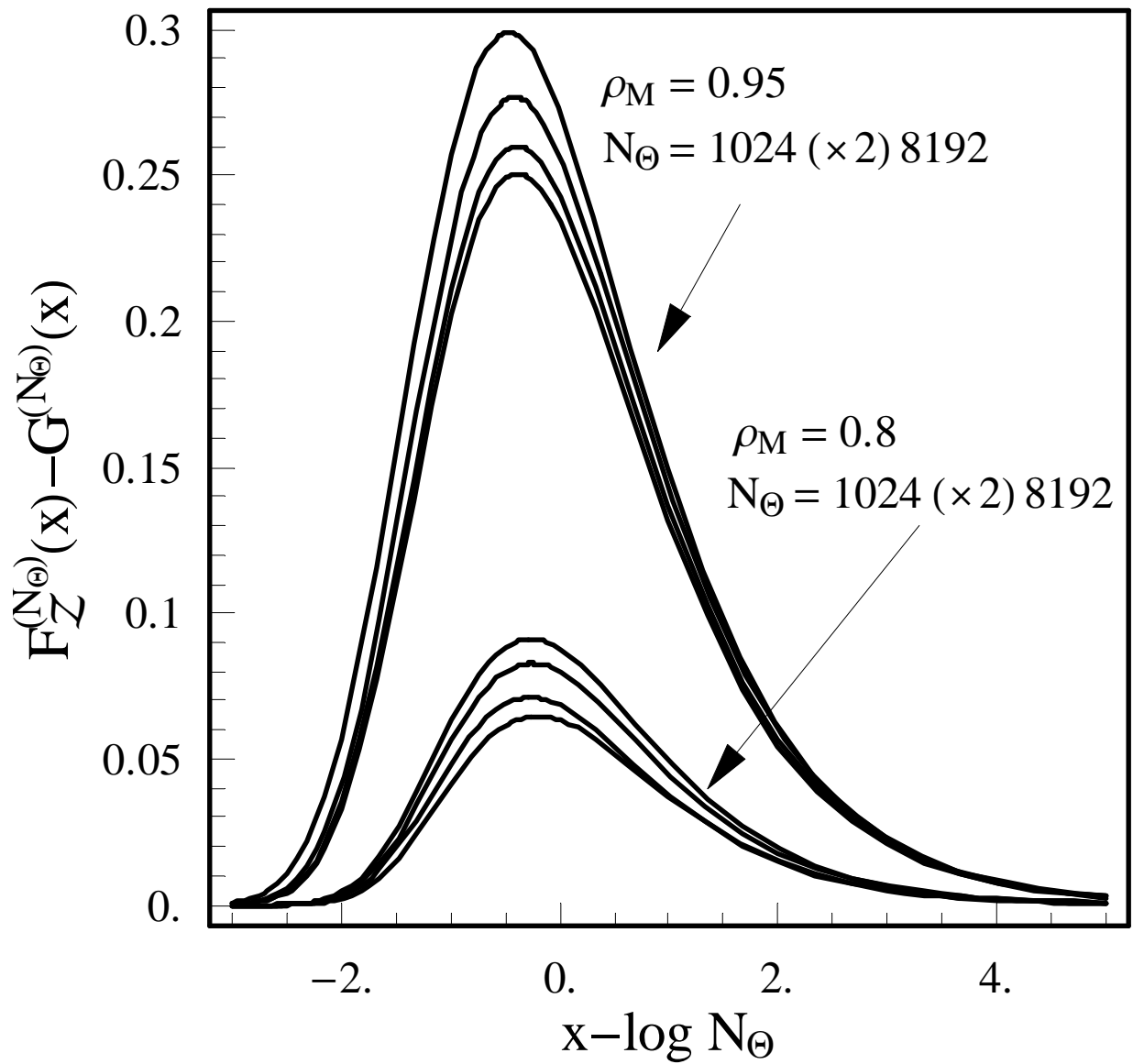
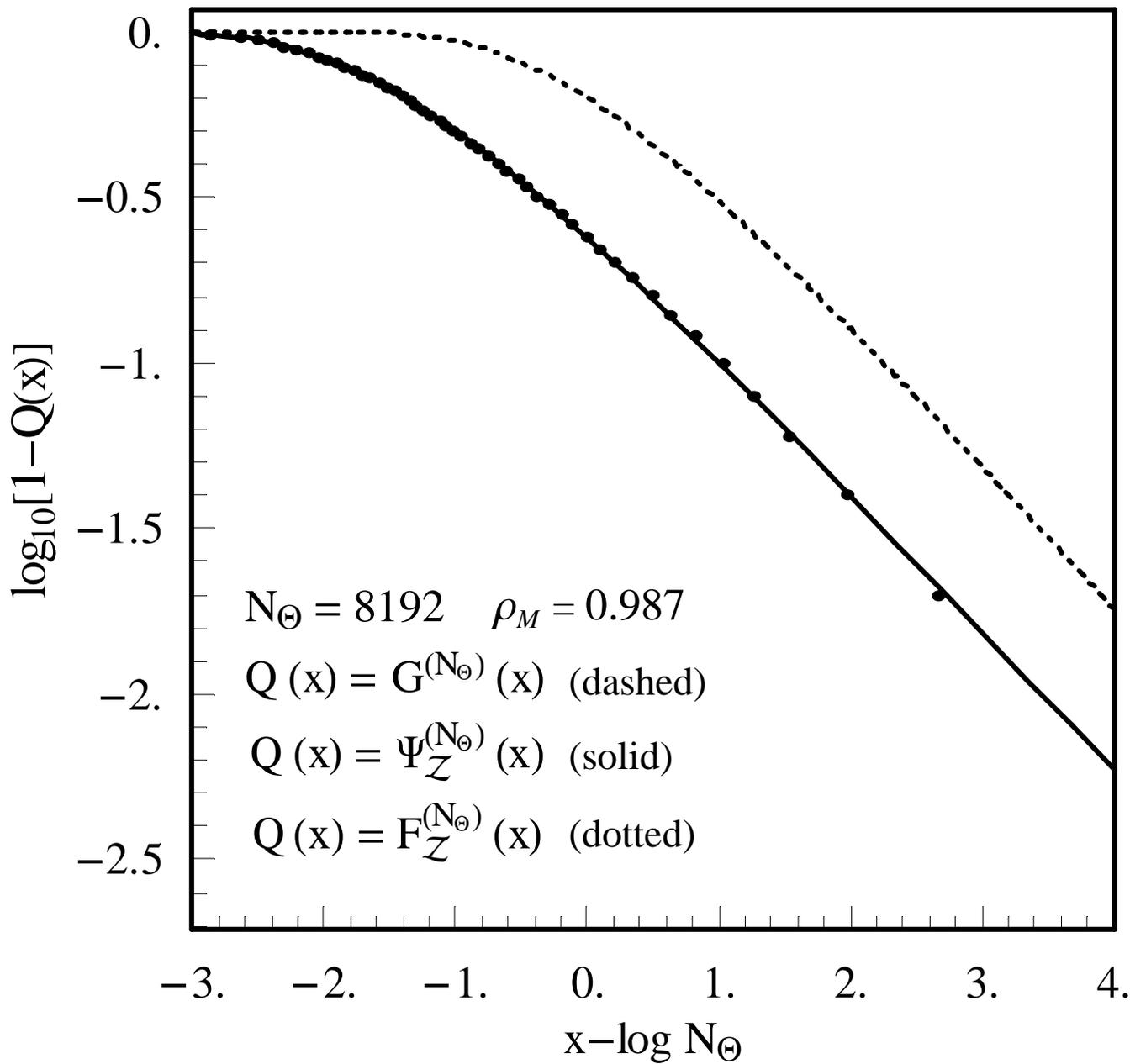
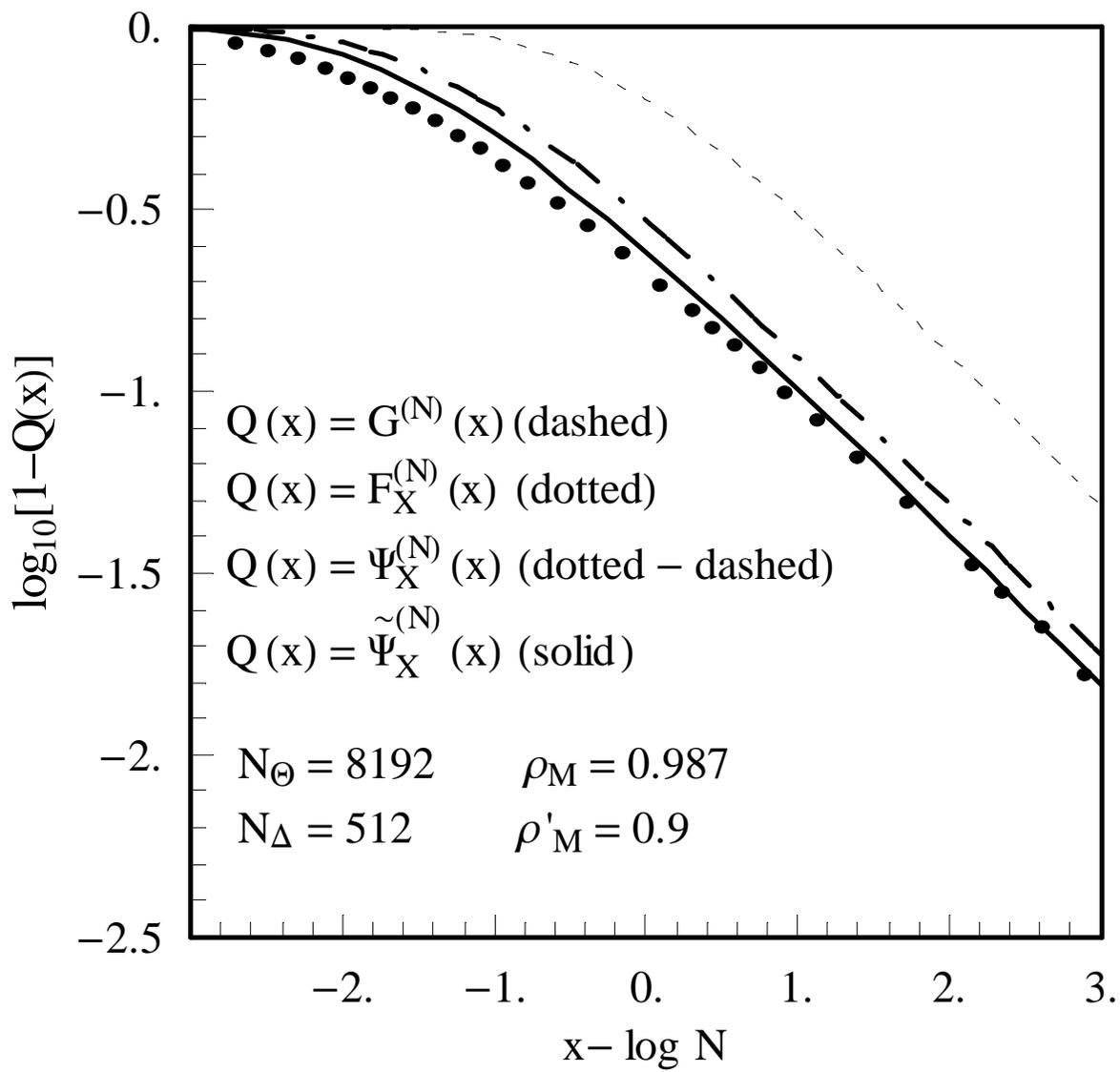
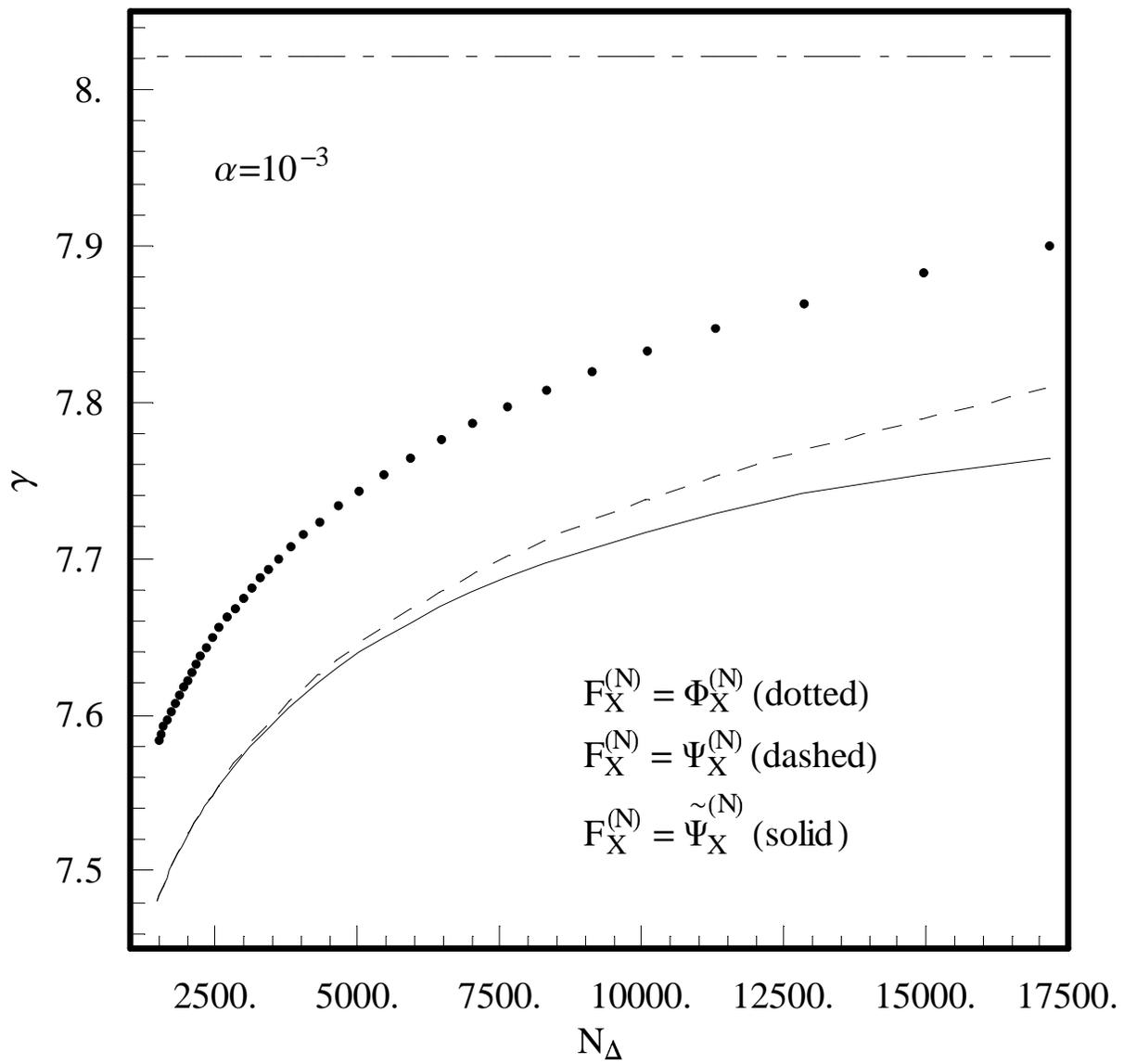
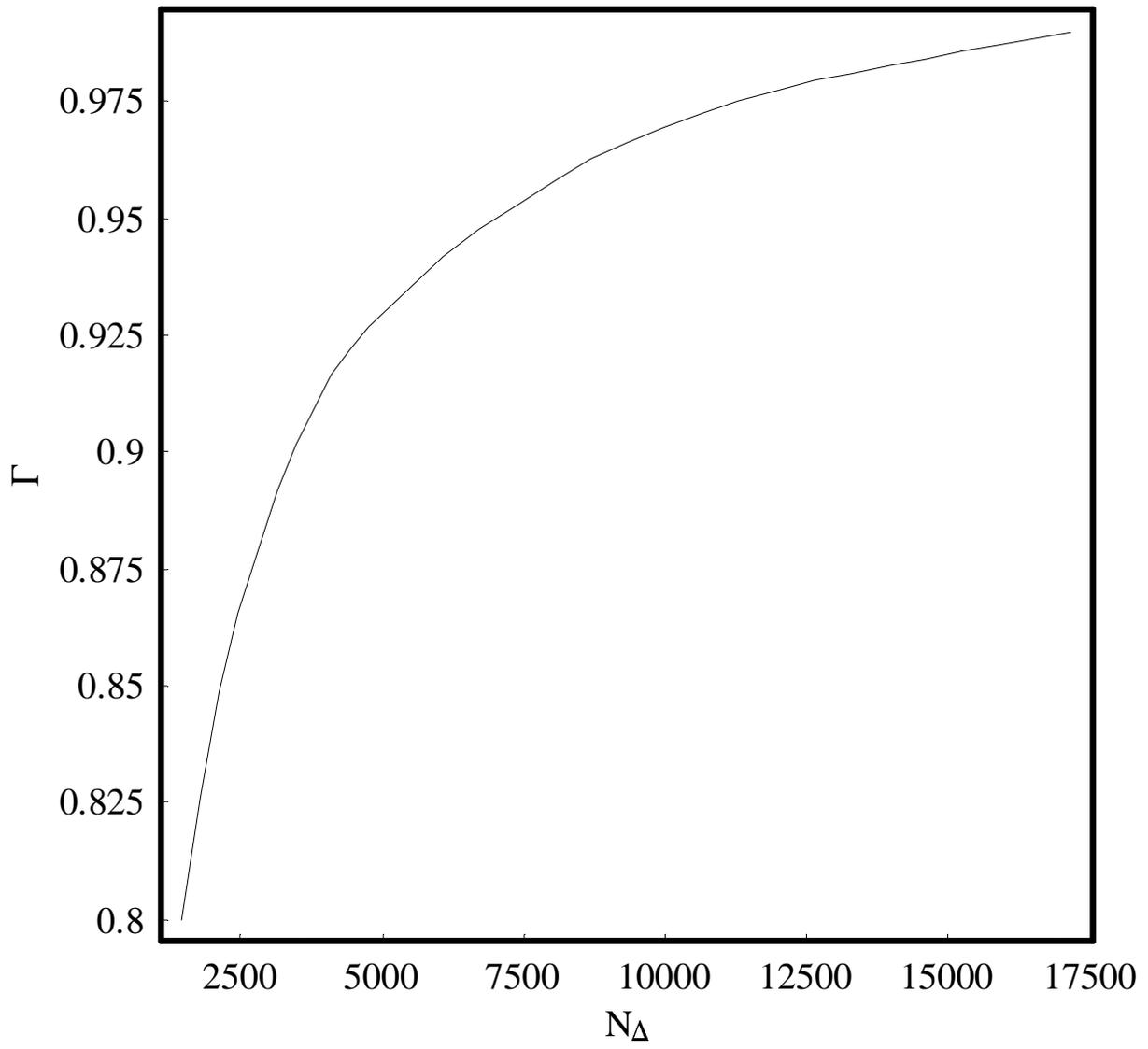
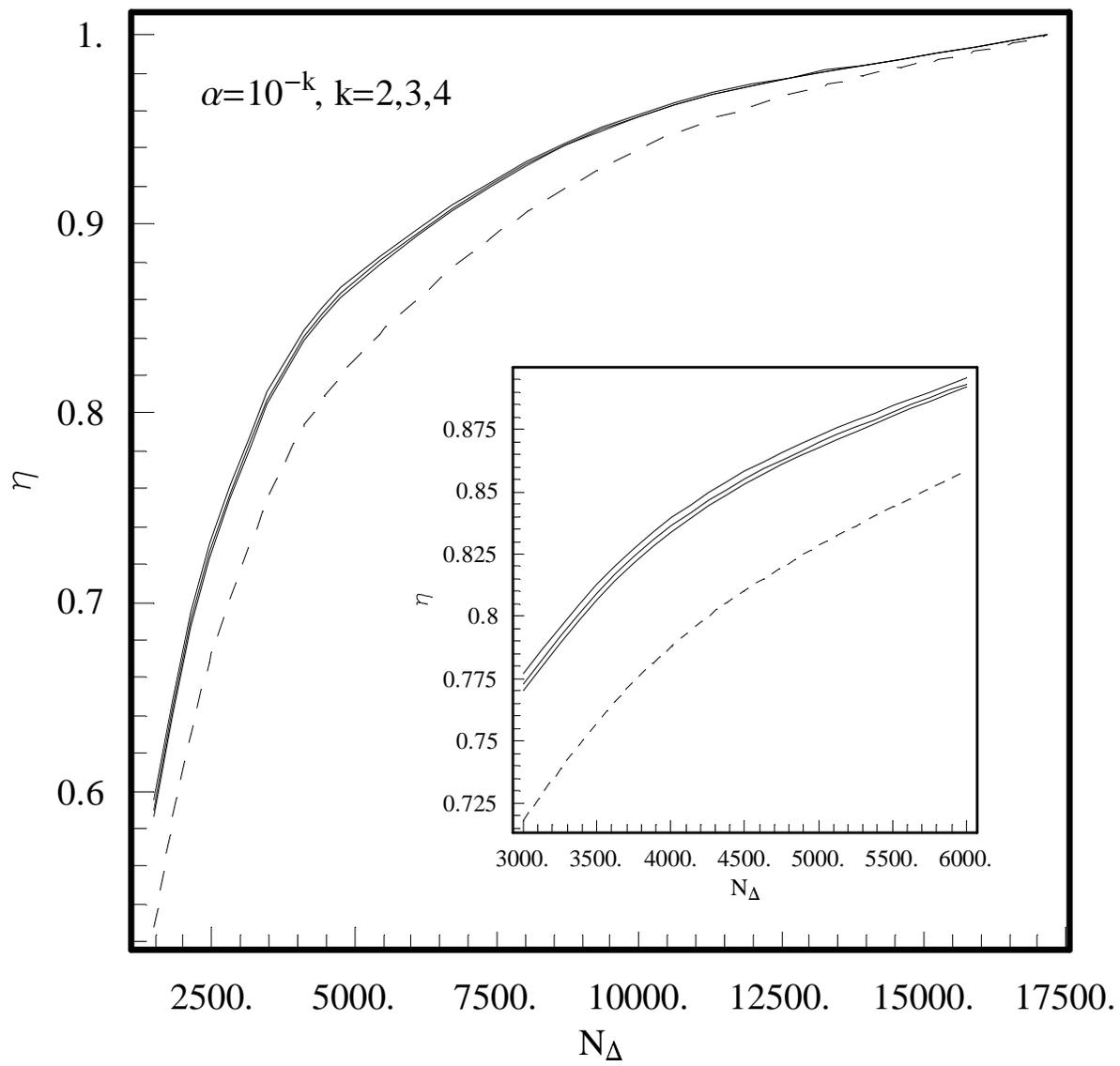
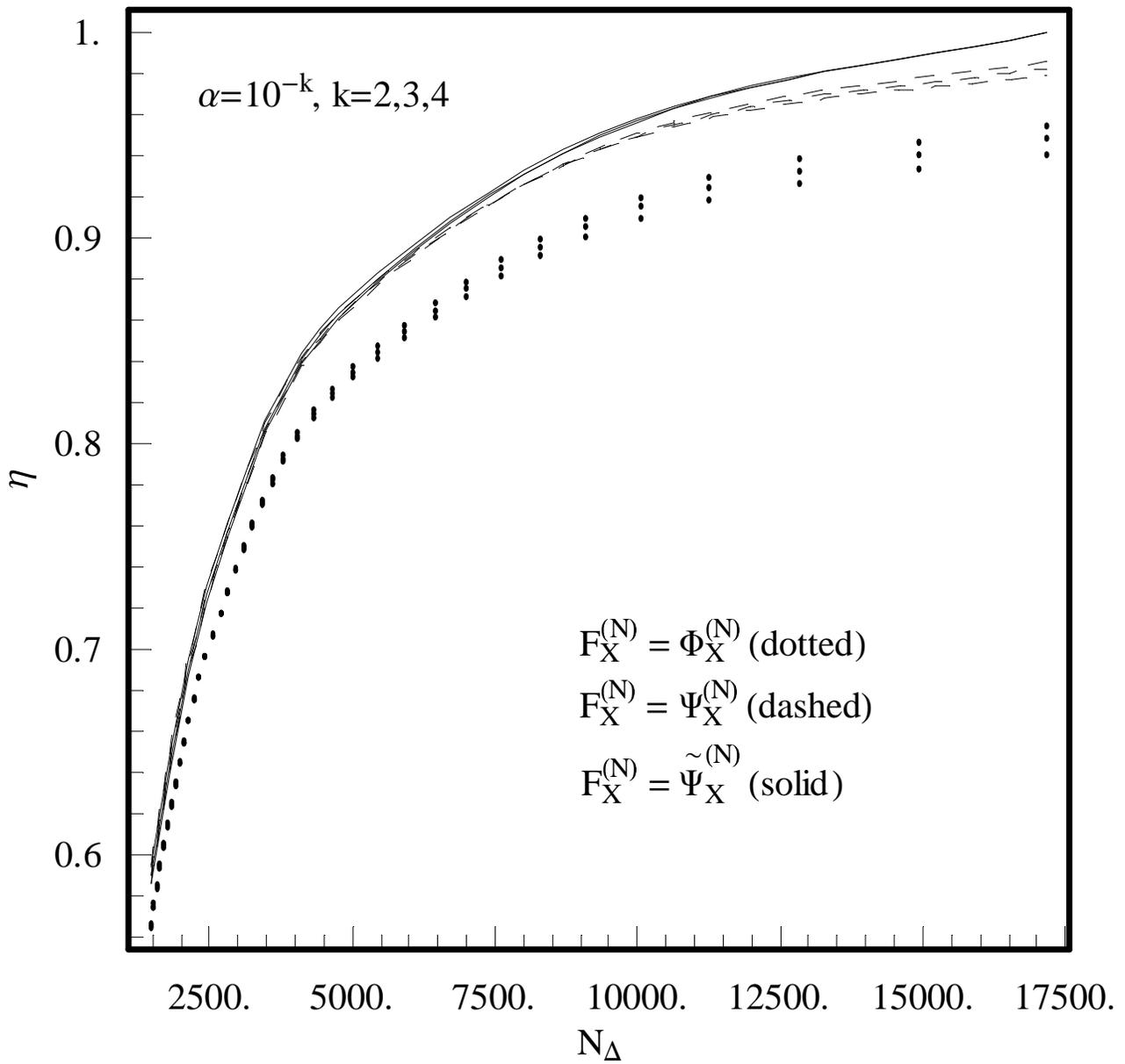
\end{document}